\documentclass[11pt]{article}

\newlength{\vshift}
\newlength{\hshift}
\usepackage{amssymb,amsopn}

\renewcommand{\theequation}{\thesection.\arabic{equation}}
\newcommand{\initiate}{\setcounter{equation}{0}}

\def\nn{\nonumber}

\def\be{\beta}
\def\a{\alpha}

\def\g{\gamma}

\def\ds{\stackrel{\star}{,}}

\def\t{\tilde}

\def\tr{{\rm Tr}}
\def\slash{{\rlap /}}

\def\kbar{{\mathchar'26\mkern-9muk}}

\def\nn{\nonumber}
\def\be{\begin{equation}}             \def\ee{\end{equation}}
\def\ba#1{\begin{array}{#1}}          \def\ea{\end{array}}
\def\bea{\begin{eqnarray} }           \def\eea{\end{eqnarray} }
\def\beann{\begin{eqnarray*} }        \def\eeann{\end{eqnarray*} }
\def\beal{\begin{eqalign}}            \def\eeal{\end{eqalign}}
             
\def\bsubeq{\begin{subequations}}     \def\esubeq{\end{subequations}}
\def\bitem{\begin{itemize}}           \def\eitem{\end{itemize}}
                    
\def\pa{\partial}
\def\a{\alpha}
\def\b{\beta}

\def\d{\delta}
\def\e{\epsilon}                
\def\g{\gamma}

\def\l{\lambda}
\def\m{\mu}
\def\n{\nu}

\def\r{\rho}                    
\def\s{\sigma}                  
\def\t{\tau}

\def\G{\Gamma}

\def\L{\Lambda}
\def\O{\Omega}

\usepackage{mathtools}
\usepackage{amsmath}
\usepackage{color}

\def\nea{(\nabla_{\alpha}e_{\mu}^{a})}
\def\neb{(\nabla_{\beta}e_{\nu}^{b})}

\def\mass{\left(m-\frac{2}{l}\right)}

\makeatletter
\@addtoreset{equation}{enumi}
\makeatother

\usepackage{slashed}
\newcommand\reallywidehat[1]{%
\savestack{\tmpbox}{\stretchto{%
  \scaleto{%
    \scalerel*[\widthof{\ensuremath{#1}}]{\kern-.6pt\bigwedge\kern-.6pt}%
    {\rule[-\textheight/2]{1ex}{\textheight}}
  }{\textheight}%
}{0.5ex}}%
\stackon[1pt]{#1}{\tmpbox}%
}
\begin{document}
\vspace{1.5cm}
\begin{titlepage}

\begin{center}

{\LARGE{\bf Noncommutative Electrodynamics from $SO(2,3)_\star$ Model of Noncommutative Gravity}}

\vspace*{1.5cm}

{{\bf Marija Dimitrijevi\' c \' Ciri\'c, Dragoljub Go\v canin,\\ 
Nikola Konjik and Voja Radovanovi\'c}}

\vspace*{1cm}

University of Belgrade, Faculty of Physics \\
Studentski trg 12, 11000 Beograd, Serbia \\[1em]

\end{center}

\vspace*{2cm}

\date{\today}

\begin{abstract}
In our previous work we have constructed a model of noncommutative (NC) gravity based on $SO(2,3)_\star$ gauge symmetry. In this paper we extend the model by adding matter fields: fermions and a $U(1)$ gauge field. Using the enveloping algebra approach and the Seiberg-Witten map we construct actions for these matter fields and expand the actions up to first order in the noncommutativity (deformation) parameter. Unlike in the case of pure NC gravity, first non-vanishing NC corrections are linear in the noncommutativity parameter. In the flat space-time limit we obtain a non-standard NC Electrodynamics. Finally, we discuss effects of noncommutativity on relativistic Landau levels of an electron in a constant background magnetic field and in addition we calculate the induced NC magnetic dipole moment of the electron.
\end{abstract}
\vspace*{1cm}


\vspace*{1cm}
\quad\scriptsize{eMail: dmarija@ipb.ac.rs, dgocanin@ipb.ac.rs, konjik@ipb.ac.rs, rvoja@ipb.ac.rs}
\vfill

\end{titlepage}

\setcounter{page}{1}
\newcommand{\Section}[1]{\setcounter{equation}{0}\section{#1}}
\renewcommand{\theequation}{\arabic{section}.\arabic{equation}}

\section{Introduction}
In the past several decades there has been a considerable effort to develop a theory that would resolve singularity issues that plague the physics of curved space-time and enable us to think beyond the concepts of Quantum Field Theory (QFT) and General Relativity (GR). Noncommutative (NC) Field Theory, as a theory of fields on NC space-time, offers a new perspective to the problem. In NC Field Theory, space-time coordinates are proclaimed to be mutually incompatible. Analogously to the Heisenberg's uncertainty relations for a conjugate coordinate-momentum pair of a particle, there exist a lower bound for the product of uncertainties $\Delta x^{\m}\Delta x^{\n}$ for a pair of two different coordinates. In order to capture this "pointlessness" of space-time, an abstract algebra of NC coordinates is introduced as a deformation of the ordinary commutative space-time structure. These NC coordinates, denoted by $\hat{x}^{\m}$, satisfy some non-trivial commutation relations, and so, it is no longer the case that $\hat{x}^{\m}\hat{x}^{\n}=\hat{x}^{\n}\hat{x}^{\m}$. The simplest case of noncommutativity is the so-called canonical noncommutativity, defined by  
\be [\hat{x}^\m, \hat{x}^\n]=i\theta^{\m\n}\, ,\label{can-kom-rel}\ee
where $\theta^{\m\n}$ are components of a constant antisymmetric matrix. Abandoning the concept of commutative (classical, smooth) space-time leads to various new physical effects, such as UV/IR mixing \cite{UVIR}, new interactions in NC deformations of Standard Model \cite{NCSM}, fuzzy geometry \cite{fuzzy} and many others. 

\noindent Instead of deforming an abstract algebra of coordinates one can take an alternative, but equivalent, approach in which noncommutativity appears in the form of NC products of functions (NC fields) of commutative coordinates. These products are called star products ($\star$-products). Specifically, the canonical noncommutativity (\ref{can-kom-rel}) is based on the NC Moyal-Weyl $\star$-product,  
\begin{equation}\begin{split}
\label{moyal}  
(\hat{f}\star\hat{g})(x) =&
      e^{\frac{i}{2}\,\theta^{\a\b}\frac{\pa}{\pa x^\a}\frac{\pa}{ \pa
      y^\b}} f (x) g (y)|_{y\to x}\\
      =& f(x)\cdot g(x) +\frac{i}{2}\theta^{\alpha\beta}\partial_\alpha f(x) \partial_\beta g(x) + {\cal O} (\theta^2) \, .
\end{split}\end{equation}
The first term in the expansion of the exponential is the ordinary point-wise multiplication of functions. The constant deformation parameters $\theta^{\alpha\beta}$ have dimensions of $(\textit{length})^{2}$ and are assumed to be small\footnote{To be more precise, the Moyal-Weyl 
$\star$-product should be written as
$$
(\hat{f}\star \hat{g})  (x) =
      e^{\frac{i}{2}\,\kbar\theta^{\a\b}\frac{\pa}{\pa x^\a}\frac{\pa}{ \pa
      y^\b}} f (x) g (y)|_{y\to x}\, ,
$$      
with the small deformation parameter $\kbar$ and arbitrary constant 
antisymmetric matrix elements $\theta^{\a\b}$. In the usual notation $\kbar$ is 
absorbed in the matrix elements $\theta^{\a\b}$ and these are called 
small deformation parameters.
}. They are considered to be fundamental constants, like the Planck length or the speed of light. However, it is also possible to treat NC deformation parameters as dynamical fields, see \cite{DynNC}.

\noindent Apart from being an interesting subject by themselves, NC Field Theories had begun to gain interest when they were recognised as a low-energy limit of a more fundamental theory of open stings. In the work of Seiberg and Witten \cite{SWMapEnvAlgebra} it is argued that coordinate functions of the endpoints of an open string constrained to a D-brane in the presence of a constant Neveu-Schwarz B-field, with $B\sim 1/\theta$, satisfy the constant noncommutativity algebra (\ref{can-kom-rel}). Thus, any kind of evidence of noncommutativity of space-time, interpreted as a low-energy effect of the String Theory, would be of great significance.

\noindent Formulation of pure gravity in a noncommutative space-time is a very interesting and a very important problem and it has been investigated using different approaches \cite{NCGrSvi}. In our previous work, we have established a model of pure NC gravity by treating it as a gauge theory of $SO(2,3)_{\star}$ group \cite{Us-16, MiSO23Razno}. The first non-vanishing NC correction to GR is at the second order in the NC parameter $\theta^{\alpha\beta}$ and it pertains even when metric is flat thus leading to a non-trivial NC deformation of Minkowski space \cite{Us-16}. Studying the deformed metric of Minkowski space, it became clear that by introducing constant noncommutativity of space-time, we are implicitly working in a preferred coordinate system - the Fermi inertial coordinates. 

\noindent We have to remember that if one aims to explain problems such as dark energy, inflation or construct a NC deformation of supergravity, one needs to include matter field and their couplings with the gravitational field. As a first step towards including matter fields in the $SO(2,3)_\star$ NC gravity model, we introduced non-interacting Dirac fermions in curved space-time. In \cite{VojaGocanin2017} we have found the first order NC correction to the Dirac action in curved space-time with various new interaction terms. As in the case of pure gravity, noncommutativity pertains in the flat space-time limit and causes linear NC deformation of electron's dispersion relation. Also, the NC-deformed energy levels are helicity-dependent, meaning that NC space-time behaves as birefringent medium for electrons propagating in it.

\noindent Following the same line of investigation as in \cite{VojaGocanin2017} we seek to obtain a complete theory of Noncommutative Electrodynamics in $SO(2,3)_{\star}$ model of noncommutative gravity that would be capable of providing some tangible predictions concerning the potentially observable physical effects of space-time noncommutativity. To include interacting Dirac fermions, we upgrade the gauge group to $SO(2,3)_{\star}\otimes U(1)_\star$ thus introducing electromagnetic field in the framework. New interaction terms that emerge enable us to study NC Electrodynamics both in curved and flat space-time. In this paper we analyse some phenomenological consequences of this new model of NC Electrodynamics in the flat space-time limit.

\noindent The paper is organized as follows. To begin with, in the following section, we discuss coupling of matter fields with gravity in the first order formalism. In particular, we are interested in incorporating matter fields in the commutative (undeformed) $SO(2,3)$ gauge theory of gravity. In Section 3 we generalize results from Section 2 to the NC $SO(2,3)_\star$ gravity. Using the Seiberg-Witten map we construct actions and calculate equations of motion for the NC $U(1)_\star$ gauge field and the NC Dirac field. Unlike in \cite{VojaGocanin2017}, fermions are now coupled with the NC $U(1)$ gauge field and the NC gravity. Finally, in Section 5 we formulate NC Electrodynamics induced by NC $SO(2,3)_\star$ gravity. In the limit of flat space-time, we discuss the equation of motion of an electron in the background electromagnetic field. Especially, we find NC corrections to its energy levels in constant magnetic field, that is, NC corrections to the relativistic Landau levels. In addition, we derived the induced NC magnetic dipole moment of an electron. We end the paper with some discussion of the obtained results and proposals for future research.

\initiate 

\section{Matter fields in $SO(2,3)$ gauge theory of gravity}

It is well known that in the first order formalism (gauge theories of gravity) fermions couple naturally to the gravitational field. On the other hand, to couple gauge fields to the gravitational field one normally requires the existence of Hodge dual operation. The definition of Hodge dual operation in the presence of gravitational field (curved space-time) requires the existence of a metric tensor, which means working in the second order formalism. This difference
becomes even more evident in the $SO(2,3)$ model of gravity. Namely, in this model the basic variable is a $SO(2,3)$ gauge field, which splits into the $SO(1,3)$ spin-connection and vierbeins only after the gauge fixing (symmetry breaking) \cite{Us-16}. In this section we discuss the construction of actions involving gauge field and Dirac spinor field in the $SO(2,3)$ gravity model.  

\noindent Let us briefly review the basics of $SO(2,3)$ gravity theory. We assume that space-time has the structure of $D=4$ dimensional Minkowski  space. The gauge field takes values in the $SO(2,3)$ algebra, $\omega_\mu = \frac{1}{2}\omega_\mu^{\;\;AB}M_{AB}$. The
generators of the $SO(2,3)$ group, $M_{AB}$, fulfil
\begin{equation}
[M_{AB},M_{CD}]=i(\eta_{AD}M_{BC}+\eta_{BC}M_{AD}-\eta_{AC}M_{BD}-\eta_{BD}M_{AC
})\, .
\label{AdSalgebra}
\end{equation}
The $5 D$ metric is $\eta_{AB}={\rm diag}(+,-,-,-,+)$. Group indices $A,B,\dots$ take
values $0,1,2,3,5$, while indices $a,b,\dots$ take values $0,1,2,3$. A representation
of the algebra (\ref{AdSalgebra}) is given by
\begin{equation}
M_{ab} =\frac{i}{4}[\gamma_a,\gamma_b]=\frac12\sigma_{ab}\, , \quad
M_{5a} =\frac{1}{2}\gamma_a\, , \label{Maba5}
\end{equation}
where $\gamma_a$ are four dimensional Dirac gamma matrices. Then the gauge
potential $\omega_\mu^{\;\;AB}$ decomposes into
$\omega_\mu^{\;\;ab}$ and $\omega^{\;\;a5}_\mu=\frac{1}{l}e_\mu^a$ :
\begin{equation}
\omega_\mu =
\frac{1}{2}\omega_\mu^{\;\;AB}M_{AB}=\frac{1}{4}\omega_\mu^{\;\;ab}\sigma_{ab}-
\frac{1}{2l}e_\mu^{a}\gamma_a \, .\label{GaugePotAdsDecomp}
\end{equation}
The field strength tensor is defined in the usual way by
\begin{eqnarray}
F_{\mu\nu} &=&
\partial_\mu\omega_\nu-\partial_\nu\omega_\mu-i[\omega_\mu,\omega_\nu]=
\frac{1}{2} F_{\mu\nu}^{\;\;\;\;AB }M_{AB}
\nonumber\\
&=& \Big( R_{\mu\nu}^{\;\;\ ab}-\frac{1}{l^2}(e_\mu^ae_\nu^b-e_\mu^be_\nu^a)\Big)
\frac{\sigma_{ab}}{4} - F_{\mu\nu}^{\;\;\ a5}\frac{\gamma_a}{2}\, , \label{FabFa5}
\end{eqnarray}
with
\begin{equation}
R_{\mu\nu}^{\;\;\ ab} =
\partial_\mu\omega_\nu^{\;\;ab}-\partial_\nu\omega_\mu^{\;\;ab}
+ \omega_\mu^{\;\;ac}\omega_\nu^ {\;\;cb }
-\omega_\mu^{\;\;bc}\omega_\nu^{\;\;ca} \, , \quad
lF_{\mu\nu}^{\;\;\ a5} = \nabla_\mu e^a_\nu - \nabla_\nu e^a_\mu = T_{\mu\nu}^{\;\;\;\;a} \, .\label{RabTa}
\end{equation}
Equations (\ref{GaugePotAdsDecomp}), (\ref{FabFa5}) and (\ref{RabTa}) suggest that
one can identify $\omega^{\;\;ab}_\mu$
with the spin connection of the Poincar\' e gauge theory, $\omega^{\;\;a5}_\mu$ with
the vierbeins, $R^{\;\;\;ab}_{\mu\nu}$ with the curvature tensor and $lF^{\;\;\;a5}_{\mu\nu}$
with torsion. It was shown in the seventies that one can indeed make such an identification and relate AdS
gauge theory with GR. Different actions were discussed in the literature, see
\cite{stelle-west, Wilczek}. A necessary step in obtaining GR from $SO(2,3)$ gravity model is the gauge fixing, that is, the symmetry breaking from local $SO(2,3)$ down to local $SO(1,3)$. In order to break $SO(2,3)$ gauge symmetry one usually introduces an
auxiliary field $\phi=\phi^A\G_A$ \cite{stelle-west}. This field is a space-time scalar and an internal-space vector. It transforms in the adjoint representation of $SO(2,3)$ group, i.e. 
\be 
\d_{\varepsilon}\phi=i[\varepsilon, \phi] \, ,
\ee
where $\varepsilon = \frac{1}{2}\varepsilon_\mu^{\;\;AB}M_{AB}$ is an infinitesimal gauge parameter. This auxiliary field also satisfies the constraint $\phi^{A}\phi_{A}=l^{2}$. Note that this field has mass dimension $-1$.

\noindent In our previous work \cite{Us-16} we analysed the pure gravity action in the $SO(2,3)$ model and we also constructed its NC generalization. We will not repeat that discussion here. For completeness we
just write the gravity action before and after the gauge fixing: 
\begin{eqnarray}
S&=&c_1S_1+c_2S_2+c_3S_3 \, ,\nn\\
S_1 &=& \frac{il}{64\pi G_N}\tr \int{\rm d}^4x\; \epsilon^{\mu\nu\rho\sigma}
F_{\mu\nu} F_{\rho\sigma}\phi  \, ,\label{KomDejstvo_S_1}\\
S_2 &=& \frac{1}{128 \pi G_{N}l}\tr\int {\rm d}^{4}x\;\epsilon^{\mu \nu \rho 
\sigma}F_{\mu 
\nu}D_{\rho}\phi D_{\sigma}\phi\phi+h.c. \, , \label{KomDejstvo_S_2}\\
S_3 &=& -\frac{i}{128 \pi G_{N}l}\tr\int {\rm d}^{4}x\;\epsilon^{\mu \nu \rho 
\sigma}D_{\m}\phi D_{\n}\phi D_{\rho}\phi D_{\sigma}\phi\phi \, ,\label{KomDejstvo_S_3} 
\end{eqnarray}
where covariant derivative in the adjoint representation is given by
\begin{equation}
D_\mu\phi = \partial_\mu\phi -i[\omega_\mu, \phi]\, . \label{cov-dev-adj}
\end{equation}
We break the $SO(2,3)$ gauge symmetry by fixing the auxiliary field, specifically, we set $\phi^a=0$ and $\phi^5=l$, and obtain
\begin{eqnarray} 
S&=&-\frac{1}{16\pi G_{N}}\int 
{\rm d}^{4}x\; \Big( c_1\frac{l^2}{16}\epsilon^{\m\n\r\s}
\epsilon_{abcd}R_{\m\n}^{\;\;\ ab}R_{\r\s}^{\;\;\ cd} \nn\\
&& +\sqrt{-g}\big( (c_1 + c_2)R -\frac{6}{l^2}(c_1+ 2c_2 + 2c_3)\big)
\Big)\, . \label{KomDejstvo} 
\end{eqnarray}
For generality, we introduced three dimensionless constants that are a priori undetermined and can be fixed by some consistency conditions. The Einstein-Hilbert term requires $c_1+c_2=1$, while the absence of the cosmological constant is provided with $c_1+ 2c_2 + 2c_3=0$. Applying both constraints leaves one free parameter. \\[0.2cm]
Now we study the coupling of matter fields with gravity in the framework of the $SO(2,3)$ model.

\subsection{$U(1)$ gauge field}

In order to include the electromagnetic interaction of electrons in our framework, we upgrade the original $SO(2,3)$ gauge group to $SO(2,3)\otimes U(1)$. Gauge potential for the whole gauge group, the master potential, consists of two independent parts:
\be
\Omega_{\m}=\omega_{\m}+A_{\m}\, .\label{MasterGaugePot}
\ee
The first part is the already mentioned $SO(2,3)$ gauge potential (\ref{GaugePotAdsDecomp}) and the second part, $A_\m$, is the electromagnetic potential. 

\noindent The field strength associated with the master gauge potential $\Omega_\m$ is
\be
\mathbb{F}_{\m\n}=\partial_\m\Omega_\n-\partial_\n\Omega_\m-i[\Omega_\m,\Omega_\n]\, ,
\ee
and it can be decomposed as 
\be
\mathbb{F}_{\m\n}=F_{\m\n}+\mathcal{F}_{\m\n}\, ,\label{MasterFmunu}
\ee 
where the gravity field strength $F_{\mu\nu}$ is given by (\ref{FabFa5}) and
\begin{equation}
\mathcal{F}_{\m\n} = \partial_\mu A_\nu - \partial_\nu A_\mu \label{FU1}
\end{equation}
is the $U(1)$ field strength tensor, i.e. the electromagnetic field.

\noindent Following the approach of \cite{PLGaugeGr} we define a $SO(2,3)$ invariant action for the $U(1)$
gauge field as follows:
\bea
S_{A}= -\frac{1}{16l} \tr\int {\rm d}^{4}x\;\epsilon^{\mu\nu\rho\sigma}\Big( f \mathbb{F}_{\m\n}D_\r\phi
D_\s\phi\phi+\frac{i}{3!}ffD_\m\phi D_\n\phi D_\r\phi D_\s\phi\phi \Big) + h.c. \,
.\label{ActionGaugeComm}
\eea
The action (\ref{ActionGaugeComm}) includes an additional auxiliary field
$f$ 
\begin{equation}
f=\frac{1}{2}f^{AB}M_{AB}\, , \quad \delta_\varepsilon f = i[\varepsilon, f] \label{fProperties} 
\end{equation}
with the gauge parameter 
\begin{equation}
\varepsilon = \frac{1}{2}\varepsilon^{AB}M_{AB} +\alpha \label{GaugeParamFullCommutative} 
\end{equation}
consisting of the $SO(2,3)$ and the $U(1)$ part. We see from (\ref{fProperties}) that the field $f$ transforms in the adjoint
representation of $SO(2,3)$ and it is invariant under $U(1)$, i.e. it is not charged. Its role is to
produce the canonical kinetic term for the $U(1)$ gauge field since we cannot define the Hodge dual operation without prior knowledge of the metric tensor.

\noindent The full covariant derivative of the field $\phi$ is defined as
\begin{equation}
D_\mu\phi = \partial_\mu\phi -i[\Omega_\mu, \phi] = \partial_\mu\phi -i[\omega_\mu, \phi] \, . \label{cov-dev-adjNew}
\end{equation}
Thus we see that the field $\phi$ is also neutral under the $U(1)$ gauge transformations. This
simplification is a peculiarity of the Abelian $U(1)$ group and it does not hold in a more general
case of non-Abelian Yang-Mills theory.

\noindent The action (\ref{ActionGaugeComm}) can be rewritten in a more explicit form as
\begin{align}
S_{A}=& -\frac{1}{16l}\int {\rm d}^{4}x\;
\epsilon^{\mu\nu\rho\sigma}\Big\{ \frac{1}{4}f^{AB}F_{\m\n}^{\;\;\;\;CD}(D_\r\phi)^{E}
(D_\s\phi)^{F}\phi^{G} \tr(M_{AB}M_{CD}\Gamma_{E}\Gamma_{F}\Gamma_{G}) \nn\\
&+\frac{i}{24} f^{AB}f^{CD}(D_\m\phi)^{E}(D_\n\phi)^{F}(D_\r\phi)^{
G}(D_\s\phi)^{H}\phi^{R}\tr(M_{AB}M_{CD}\Gamma_{E}\Gamma_{F}\Gamma_{G}\Gamma_{H}\Gamma_{R})\nn\\
&+\frac{1}{2}f^{AB}\mathcal{F}_{\m\n}(D_\r\phi)^{E}(D_\s\phi)^{F}\phi^{G} \tr(M_{AB}\Gamma_{E}
\Gamma_{F}\Gamma_{G})\;\Big\} +h.c. \, .
\end{align}
After calculating traces (see Appendix A) we obtain
\begin{align}
S_{A}= & -\frac{i}{32l}\int {\rm d}^{4}x\;
\epsilon^{\mu\nu\rho\sigma}\Big\{ f^{AB}F_{\m\n}^{\;\;\;\;CD}(D_\r\phi)^{E}(D_\s\phi)^{F}\phi^{G
}(\eta_{FG}\epsilon_{ABCDE}
+2\eta_{AD}\epsilon_{BCEFG}) \nn\\
& -2i f^{AB}\mathcal{F}_{\m\n}(D_\r\phi)^{E}(D_\s\phi)^{F}
\phi^{G}\epsilon_{ABEFG} \nn\\
& -\frac{i}{6  }f^{AB}f_{AB}(D_\m\phi)^{E}(D_\n\phi)^{F}(D_\r\phi)^{G}
(D_\s\phi)^{H}\phi^{R}\epsilon_{EFGHR} \Big\} +h.c. \, . \label{class-Sc-after traces}  
\end{align}
The first term in (\ref{class-Sc-after traces}) is purely imaginary and it will vanish because we have defined $S_{A}$ to be real by adding corresponding hermitian conjugate terms. Thus, after the gauge fixing, when $(D_\m\phi)^{a}=e_\m^a$ and $(D_\m\phi)^{5}=0$, the action reduces to  
\begin{align}
S_{A}& = -\frac{1}{8}\int {\rm d}^{4}x\;\epsilon^{\m\n\r\s}\Big\{ f^{ab}\mathcal{F}_{\m\n}\epsilon_{abef}e_\r^e
e_\s^f +\frac{1}{12}f^{AB}f_{AB}\;\epsilon_{efgh}e_\m^e e_\n^f
e_\r^g e_\s^h \Big\}\nn\\
& = \frac{1}{2}\int {\rm d}^{4}x\;e\; \Big( f^{ab}e_a^\m e_b^\n\mathcal{F}_{\m\n} +\frac{1}{2}(f^{ab}f_{ab}+2f^{a5}f_{a}^{\;\;5})\Big) \, . \label{SA}
\end{align}
with the vierbein determinant $e=det(e_{\m}^{a})=\sqrt{-g}$\, .

\noindent Equations of motion (EOMs) for the components of the auxiliary field $f$ are
\begin{align}
f_{a5}=0 \, , \;\;\;\;f_{ab}=-e_a^\m e_b^\n\mathcal{F_{\m\n}}\, . \label{EOMs-0th order}
\end{align}
Using these EOMs we can eliminate the auxiliary field in the action (\ref{SA}). This leaves us with the well known action for pure $U(1)$ gauge field in curved space-time:
\be
S_{A}= -\frac{1}{4}\int {\rm d}^{4}x\;e\;g^{\mu\rho}g^{\nu\sigma}\mathcal{F}_{\m\n}\mathcal{F}_{\rho\sigma}\, .
\ee 


\subsection{Dirac fermions}

The Dirac spinor field $\psi$ transforms in the fundamental 
representation of $SO(2,3)\otimes U(1)$ gauge group, i.e.
\be 
\d_{\varepsilon} \psi=i\varepsilon\psi=\frac{i}{2}\varepsilon^{AB}M_{AB}\psi + i\alpha\psi\, , \label{TransfLawPsi}
\ee
where, as in (\ref{GaugeParamFullCommutative}), $\varepsilon^{AB}$ are infinitesimal antisymmetric gauge parameters of $SO(2,3)$ gauge group and $\alpha$ is an infinitesimal gauge parameter of $U(1)$ gauge group.
The covariant derivative of the full $SO(2,3)\otimes U(1)$ gauge group in fundamental representation is given by
\begin{align}
D_\m\psi&=\partial_\m\psi-i\Omega_\m\psi=\partial_\m\psi-i(\omega_{\m}+A_{\m})\psi \nn\\
&=\widetilde{\nabla}_\m\psi+\frac{i}{2}e_\m^a\gamma_a\psi\, ,     
\end{align}    
where we introduced $\widetilde{\nabla}_\m=\nabla_\m-iA_\m$ as a covariant derivative for $SO(1,3)\otimes U(1)$ gauge group, and $\nabla_\m$ is the ordinary $SO(1,3)$ covariant derivative. In addition, we set $q=-1$ for an electron.

\noindent The fermionic action consists of two parts: the kinetic term $S_{\psi, kin}$ (which also contains the interaction) and the mass term $S_{\psi, m}$. They are given by
\be
S_{\psi, kin} = \frac{i}{12}\int {\rm d}^{4}x\;\epsilon^{\m\n\r\s}\;\Big\{\bar{\psi}D_\m\phi D_\n\phi D_\r\phi
D_\s\psi-D_\s\bar{\psi}D_\m\phi D_\n\phi D_\r\phi\psi\Big\}\, , \label{Spsi-kin}
\ee
\begin{align}
S_{\psi, m}= &\frac{i}{144}\Big(\frac{m}{l}-\frac{2}{l^{2}}\Big)\int
{\rm d}^{4}x\;\epsilon^{\m\n\r\s}\bar{\psi}\Big\{D_\m\phi D_\n\phi D_\r\phi D_\s\phi\phi \nn\\
&-D_\m\phi D_\n\phi D_\r\phi\phi D_\s\phi +D_\m\phi D_\n\phi\phi D_\r\phi D_\s\phi\Big\}\psi+h.c.\, , \label{Spsi-mass}
\end{align}
and they were investigated in detail in \cite{VojaGocanin2017}, both classically and noncommutatively, but
without any reference to the electromagnetic interaction. The only difference however, is that the covariant
derivative of the Dirac field now contains an additional term, $-iA_{\m}$, associated with the
$U(1)$ gauge group. This will give us the interaction term for the Dirac field.

\noindent After the symmetry breaking the total spinor action $S_{\psi}$, given by the sum of $(\ref{Spsi-kin})$ and $(\ref{Spsi-mass})$, reduces to
\be
S_{\psi} = S_{\psi, kin}+S_{\psi, m} = i\int {\rm d}^{4}x\;e\;\bar{\psi}\Big( e_s^\s\gamma^s(\nabla_\s-i A_\s)-m\Big)\psi \, .
\ee
This is the familiar action for a $U(1)$ charged Dirac fermion in curved space-time.

\section{NC matter fields}

Let us now generalize the setup from the previous section and define actions for the NC $U(1)_\star$ gauge field and the NC Dirac fermion.

\noindent We will work in the canonically deformed NC space-time with the NC Moyal-Weyl
$\star$-product (\ref{moyal}). However, our construction can straightforwardly be generalized to an arbitrary noncommutative space-time coming form an Abelian twist deformation. 

\noindent To establish the NC field theory with $SO(2,3)_{\star}\otimes U(1)_{\star}$ gauge group, we introduce the NC spinor field $\widehat{\psi}$ and the NC gauge potential $\widehat{\O}_\m$. The corresponding NC field strength tensor is defined as
\be 
\widehat{\mathbb{F}}_{\m\n}=\pa_\m\widehat{\O}_\n-\pa_\n\widehat{\O}_\m
-i[\widehat{\O}_\m\ds\widehat{\O}_\n] \, . \label{nckrivina}
\ee

\noindent The covariant derivatives of the NC spinor $\widehat{\psi}$ and the auxiliary field $\widehat{\phi}$ are defined by 
\bea
D_\m\widehat{\psi}&=&\pa_\m \widehat{\psi}-i\widehat{\O}_\m 
\star\widehat{\psi}\, , \label{covdevpsi} \\
D_\m\widehat{\phi}&=&\pa_\m \widehat{\phi}-i[\widehat{\O}_\m\ds 
\widehat{\phi}]\, . \label{covdevphi}
\eea 
The fields $\widehat\psi$ and $\widehat{\phi}$, along with their covariant derivatives (\ref{covdevpsi}) and (\ref{covdevphi}), transform in the fundamental and adjoint representation, respectively, under NC infinitesimal gauge transformations, i.e.  
\begin{align}
\delta^\star_\varepsilon \widehat {\psi}&= i \widehat{\Lambda}_\varepsilon 
\star{\widehat
\psi} \ ,\;\;\;\;\delta^\star_\varepsilon D_\m \widehat{\psi}=i\widehat{\L}_\varepsilon\star 
D_\m\widehat{\psi} \, , \nn\\
\delta^\star_\varepsilon {\widehat \phi}&= i[\widehat{\Lambda}_\varepsilon 
\ds{\widehat
\phi}] \, ,\;\;\;
\delta^\star_\varepsilon D_\m\widehat{\phi}=i[\widehat{\L}_\varepsilon\ds 
D_\m\widehat{\phi}] \, . 
\label{inf-transf}\end{align}
The transformation laws for NC gauge potential and field strength are  
\begin{equation}\label{NCTransfLawOmegaF}\begin{split}
\delta^\star_\varepsilon \widehat {\O}_{\m} =& 
\pa_\m\widehat{\L}_\varepsilon-i[\widehat{\O}_\m\ds\widehat{\L}_\varepsilon] \, , \\ 
\delta^\star_\varepsilon \widehat {\mathbb{F}}_{\m\n} =& i[\widehat{\Lambda}_\varepsilon 
\ds \widehat
{\mathbb{F}}_{\m\n}] \, .
\end{split}\end{equation}
We see that NC field strength $\widehat{\mathbb{F}}_{\m\n}$ transforms in the adjoint representation of the deformed gauge group $SO(2,3)_{\star}\otimes U(1)_{\star}$ just as ordinary field strength $\mathbb{F}_{\m\n}$ transforms in the adjoint representation of $SO(2,3)\otimes U(1)$. In the previous transformation rules, 
$\widehat{\Lambda}_\varepsilon$ is the NC gauge parameter of the full $SO(2,3)_{\star}\otimes U(1)_{\star}$ gauge group, which in the commutative limit reduces to (\ref{GaugeParamFullCommutative}).

\noindent The Seiberg-Witten (SW) map enables us to express NC fields in terms of the corresponding commutative fields, without introducing new degrees of freedom (new fields) in the theory. NC fields are represented as power series in the deformation parameter $\theta^{\a\b}$, with expansion coefficients built out of the commutative quantities like $\phi$, $\psi$ and $\Omega_\m$. For example:
\bea
{\widehat\O}_\m &=& \O_\mu
-\frac{1}{4}\theta^{\a\b}\{\Omega_\a, \partial_\b\Omega_\mu +
\mathbb{F}_{\b\m}\}
+ {\cal O}(\theta^2)\, , \label{NC-Omega}  \\
{\widehat \phi} &=& \phi -\frac{1}{4}\theta^{\a\b}
\{\Omega_\a,(\pa_\b + D_\b) \phi\} + {\cal O}(\theta^2) \, , \label{phi-expansion} \\
\widehat{\psi} &=& \psi-\frac14\theta^{\alpha\beta}\Omega_{\alpha}(\partial_{\beta}+D_{\beta})\psi
+ {\cal O}(\theta^2) \, , \label{psi-expansion}
\eea
where $\Omega_\mu$ is the the commutative master gauge potential (\ref{MasterGaugePot}), while $\psi$ and $\phi$ are the commutative Dirac spinor and the auxiliary field, respectively. Using the SW map, we can derive the first order 
NC corrections to the field strength, and the covariant 
derivatives of adjoint and spinor field. They are given by
\begin{equation}\label{FDexpansion}\begin{split}
\widehat{\mathbb{F}}_{\m\n} =& \mathbb{F}_{\m\n}-\frac{1}{4}\theta^{\alpha\beta}\{\Omega_{\alpha} ,
(\partial_{\beta}+D_{\beta})\mathbb{F}_{\m\n}\}
+\frac{1}{2}\theta^{\alpha\beta}\{\mathbb{F}_{\alpha\mu},\mathbb{F}_{\beta\nu}\}+{\cal 
O}(\theta^2)\, ,\\
D_{\mu}\widehat{\phi} = & D_\m\phi-\frac{1}{4}\theta^{\alpha\beta}\{\Omega_{\alpha} 
,(\partial_{\beta}+D_{\beta})
D_{\mu}\phi\}+\frac{1}{2}\theta^{\alpha\beta}\{\mathbb{F}_{\alpha\mu},D_{\beta}\phi\}+{
\cal O}(\theta^2) \, ,\\
D_\m\widehat{\psi} =& D_\m\psi-
\frac14\theta^{\a\b}\O_\a(\pa_\b 
+D_\b)D_\m\psi+\frac{1}{2}\theta^{\a\b}\mathbb{F}_{\a\m}D_\b\psi+{\cal O}(\theta^2) \, ,
\end{split}\end{equation}
with the commutative field strength tensor $\mathbb{F}_{\mu\nu}$ defined in (\ref{MasterFmunu}). All these results will be put into use in the next subsection where we turn to the NC version of the actions (\ref{ActionGaugeComm}) and (\ref{Spsi-kin}, \ref{Spsi-mass}) and calculate their perturbative expansions in powers of the deformation parameter $\theta^{\a\b}$.

\subsection{NC $U(1)$ gauge field}

To construct a NC model of $U(1)$ gauge field coupled to gravity with $SO(2,3)_{\star}\otimes U(1)_{\star}$ symmetry, we canonically deform action (\ref{ActionGaugeComm}): 
\begin{eqnarray}
\widehat{S}_{A} &=& -\frac{1}{16l}\tr\int {\rm d}^{4}x\; \epsilon^{\mu\nu\rho\sigma}\Big( \hat{f}\star
\hat{\mathbb{F}}_{\m\n}\star D_\r\hat{\phi}\star D_\s\hat{\phi}\star \hat{\phi} \nonumber\\
&& + \frac{i}{3!}\hat{f}\star\hat{f}\star D_\m\hat{\phi}\star
D_\n\hat{\phi} \star D_\r\hat{\phi}\star D_\s\hat{\phi}\star \hat{\phi}\Big)+ h.c. \, .\label{NCActionGauge}
\end{eqnarray}
In addition to the NC master gauge field $\widehat{\Omega}$ and the NC auxiliary field $\widehat{\phi}$ we also have to introduce a NC generalization of the auxiliary field $f$ defined in (\ref{fProperties}). The NC field $\widehat{f}$ transforms in the adjoint representation of the full NC gauge group
\begin{equation}
\delta^\star_\varepsilon {\widehat f} = i[\widehat{\Lambda}_\varepsilon 
\ds{\widehat f}] \, .\label{NCTransfLawf} 
\end{equation}
The transformation laws (\ref{NCTransfLawOmegaF}) and (\ref{NCTransfLawf}) ensure that the action (\ref{NCActionGauge}) is invariant under the $SO(2,3)_{\star}\otimes U(1)_{\star}$ NC gauge transformations.

\noindent The general rule for calculating first order NC correction to the $\star$-product of two fields states that 
\begin{equation} 
\left(\widehat{A}\star\widehat{B}\right)^{(1)}
=\widehat{A}^{(1)}B+A\widehat{B}^{(1)}
+\frac{i}{2}\theta^{\alpha\beta}\partial_\a A \partial_\b B \, . \label{rule}
\end{equation}
If both of these two fields transform in the adjoint representation, one can show \cite{PLM-13} that the following statement holds,
\begin{align}
\left(\widehat{A}\star\widehat{B}\right)^{(1)}=
 &-\frac{1}{4}\theta^{\a\b}\{\Omega_\a,(\partial_\b+D_\b)AB\}
 +\frac{i}{2}\theta^{\a\b} D_\a A D_\b B \nn \\
 &+cov(\widehat{A}^{(1)})B+Acov(\widehat{B}^{(1)}) \, , \label{rule1}
\end{align}    
where $cov(\widehat{A}^{(1)})$ is the covariant part of $A's$ first order NC correction, and
$cov(\widehat{B}^{(1)})$, 
the covariant part of $B's$ first order NC correction. By using this specialised rule (\ref{rule1})
we significantly reduce the amount of calculation and immediately obtain the covariantized result,   
which is not the case with the general rule (\ref{rule}). After some simplification, including a few partial integrations, the first order NC correction to $\widehat{S}_{A}$ is given by
\begin{align}
\widehat{S}^{(1)}_{A} =& \widehat{S}^{(1)}_{Af} + \widehat{S}^{(1)}_{Aff} \nn\\  
= & \frac{1}{32l}\theta^{\a\b}\;\tr\int {\rm d}^{4}x\;\epsilon^{\m\n\r\s}\bigg\{
\frac{1}{2}\{\mathbb{F}_{\a\b}, f\} \mathbb{F}_{\m\n}D_\r\phi D_\s\phi \phi +if D_\b \mathbb{F}_{\m\n}D_\a(D_\r\phi D_\s\phi \phi) \nn\\
& - f\{\mathbb{F}_{\a\m},\mathbb{F}_{\b\n}\}D_\r\phi D_\s\phi \phi 
-if\mathbb{F}_{\m\n}D_\a(D_\r\phi D_\s\phi)D_\b\phi \nn\\
&-if\mathbb{F}_{\m\n}(D_\a D_\r\phi)(D_\b D_\s\phi)\phi \nn - f\mathbb{F}_{\m\n}[\{\mathbb{F}_{\a\r},D_\b\phi\}, D_\s\phi]\phi\\
& +\frac{i}{3!}\bigg(\frac{1}{2}\{\mathbb{F}_{\a\b}, f^{2}\} D_\m\phi D_\n\phi D_\r\phi D_\s \phi \phi - f^{2}\{[\{\mathbb{F}_{\a\m}, D_\b\phi\}, D_\n\phi], D_\r\phi D_\s\phi\}\phi \nn\\
& - if^{2} \Big( D_\a(D_\m \phi D_\n \phi D_\r\phi D_\s\phi)D_\b\phi + D_\a(D_\m\phi D_\n\phi D_\r\phi)(D_\b D_\s\phi)\phi \label{NCSA1}\\
& +\big(D_\a(D_\m \phi D_\n \phi)(D_\b D_\r\phi)
+(D_\a D_\m\phi)(D_\b D_\n\phi)D_\r\phi\big)D_\s\phi \phi \Big)\bigg)\bigg\}+ h.c. \, .\nn
\end{align}
As we can see, all terms are manifestly $SO(2,3)\otimes U(1)$ invariant. This property is insured by the Seiberg-Witten map. The $f$-part of the obtained first order action will be denoted as $\widehat{S}^{(1)}_{Af}$ and the $f^{2}$-part as $\widehat{S}^{(1)}_{Aff}$.

\noindent After the gauge fixing by choosing $\phi^a=0$ and $\phi^5=l$, the $\widehat{S}^{(1)}_{Af}$ part becomes 
\begin{equation}
\widehat{S}^{(1)}_{Af} =\sum_{j=1}^6 \widehat{S}^{(1)}_{Af.j}\, , \label{NCSAf1}
\end{equation}
with
\begin{align}
\widehat{S}^{(1)}_{Af.1} & = -\frac{1}{8}\theta^{\a\b}\int {\rm d}^{4}x\;e\;\Big\{ 
-\frac{1}{16}\epsilon_{abcd}\epsilon^{rspt}e_{p}^{\m}e_{t}^{\n} F_{\a\b}^{\;\;\;ab}f^{cd}F_{\m\n rs}  \nn\\
& + F_{\a\b}^{\;\;\;ab}f^{c5}\big( F_{\m\n a5}\;e_{b}^{\m}e_{c}^{\n} +\frac{1}{2}F_{\m\n c5}e_{a}^{\m}e_{b}^{\n}\big) \nn\\
& +F_{\a\b}^{\;\;\;a5}f^{cd} \big( F_{\m\n d5}e_{a}^{\m}e_{c}^{\n} + \frac{1}{2}F_{\m\n a5}\;e_{c}^{\m}e_{d}^{\n}\big) \nn\\
& + \frac{1}{4}F_{\m\n}^{\;\;\;mn}\;e_{m}^{\m}e_{n}^{\n} \big( F_{\a\b}^{\;\;\;ab}f_{ab}
+2 F_{\a\b}^{\;\;\;a5}f_{a5} \big) +2\mathcal{F}_{\a\b}f^{cd}\mathcal{F}_{\m\n}e_{c}^{\m}e_{d}^{\n} \Big\} \, , 
\end{align}
\begin{align}
\widehat{S}^{(1)}_{Af.2} & =  -\frac{1}{4l}\theta^{\a\b}\int {\rm d}^{4}x\;e\; \Big\{ e_{d}^{\m}\big( f_{c}^{\;\;d}(D_\b F_{\a\m}^{\;\;\;c5}) 
+ f_{c}^{\;\;5}(D_\b F_{\a\m}^{\;\;\;dc})\big) \\
& -l(\nabla_\a e_{\r}^{r})\big( f_{c}^{\;\;d}(D_\b F_{\m\n}^{\;\;\;mc}) + f_{5}^{\;\;d}(D_\b F_{\m\n}^{\;\;\;m5})\big)
(e_{d}^{\m}e_{m}^{\n}e_{r}^{\r}+e_{r}^{\m}e_{d}^{\n}e_{m}^{\r}
+e_{m}^{\m}e_{r}^{\n}e_{d}^{\r})\Big\} \, ,\nn
\end{align}
\begin{align}
\widehat{S}^{(1)}_{Af.3} & =  \frac{1}{4}\theta^{\a\b}\int {\rm d}^{4}x\;e\;\Big\{ 
-\frac{1}{16}\epsilon_{ambn}\epsilon^{cdpt}e_{p}^{\m}e_{t}^{\n}f_{cd}F_{\a\m}^{\;\;\;am}F_{\b\n}^{\;\;\;bn}\nn\\
& + F_{\a\m}^{\;\;\;am}F_{\b\n}^{\;\;\;b5} \big( f_{a5}e_{m}^{\m}e_{b}^{\n} 
+ f_{b5}e_{a}^{\m}e_{m}^{\n}\big) +2 f^{cd}\mathcal{F}_{\a\m}\mathcal{F}_{\b\n}e_{c}^{\m}e_{d}^{\n} \\
& +\frac{1}{4}e_{c}^{\m}e_{d}^{\n} \Big( f^{cd}\big(F_{\a\m}^{\;\;\;am}F_{\b\n am} 
+ 2F_{\a\m}^{\;\;\;a5}F_{\b\n a5}\big) +4f_{m5}F_{\a\m}^{\;\;\;\;dm}F_{\b\n}^{\;\;\;c5} \Big) 
\Big\} \, , \nn
\end{align}
\begin{align}
\widehat{S}^{(1)}_{Af.4} & =  -\frac{1}{16l}\theta^{\a\b}\int {\rm d}^{4}x\;e\;\epsilon^{\m\n\r\s} \Big\{ f^{a5}F_{\m\n}^{\;\;\;bc}(\nabla_\a e_\rho^d)e^{\d} _a e^{\l} _b e^{\g}_c\big( g_{\s\b}\epsilon_{\d\l\g\t}e^{\t}_d - \epsilon_{\d\l\g\s}e_{\b d}\big)\nn\\
& + f^{ab}F_{\mu\nu}^{\;\;\;c5}(\nabla_\alpha e_\rho^d)e^{\delta} _a e^{\lambda} _b e^{\gamma} _c \big( g_{\sigma\beta}\epsilon_{\delta\lambda\gamma\tau}e^{\tau} _d - \epsilon_{\d\l\g\s}e_{\b d}\big)\nn\\
& -2 (\nabla_\alpha e_\rho^e)\epsilon_{\delta\lambda\sigma\beta}e^{\delta} _b e^{\lambda} _e\big( f^{a5}F^{\;\;\;b}_{\mu\nu\;\;a} + f^{ab}F_{\mu\nu a5}\big) \\
& + \frac{1}{2l}\epsilon_{\delta\lambda\gamma\tau}e^{\delta} _a e^{\lambda} _b e^{\gamma} _c e^{\tau} _d f^{ab}F_{\mu\nu}^{\;\;\;cd}g_{\alpha\rho}g_{\sigma\beta} + \frac{2}{l}(f^{ab}F^{\;\;\;c}_{\mu\nu\;\;a}-f^{b5}F^{\;\;\;\;c5}_{\mu\nu})g_{\alpha\rho}\epsilon_{\delta\lambda\sigma\beta}e^{\delta} _b e^{\lambda} _c \Big\} \, , \nn
\end{align}
\begin{align}
\widehat{S}^{(1)}_{Af.5} & =  \frac{1}{64}\theta^{\a\b}\int {\rm d}^{4}x\;e\;\epsilon^{\m\n\r\s}\epsilon_{\delta\lambda\gamma\tau}e^{\delta} _a e^{\lambda} _b e^{\gamma} _c e^{\tau} _d f^{ab}F_{\mu\nu}^{\;\;\;cd}\Big( (\nabla_\alpha e_{\rho}^{e})(\nabla_\beta e_{\sigma e})+\frac{1}{l^2}g_{\alpha\rho}g_{\sigma\beta}\Big) \, ,
\end{align}
\begin{align}
\widehat{S}^{(1)}_{Af.6}& = -\frac{1}{32}\theta^{\a\b}\int {\rm d}^{4}x\;e\;\epsilon^{\m\n\r\s} \Big\{ \epsilon_{\delta\lambda\beta\sigma}e^{\delta}_e e^{\lambda}_f F_{\alpha\rho}^{\;\;\;ef}(f^{ab}F_{\mu\nu ab}+2f^{a5}F_{\mu\nu a5})\nn\\
& +4 \epsilon_{\delta\lambda\beta\sigma}e^{\delta}_e e^{\lambda}_b\big( f^{ab}F_{\mu\nu a5}F_{\alpha\rho}^{\;\;\;e5} + f^{a5}F^{\;\;\;b}_{\mu\nu \;\;a}F_{\alpha\rho}^{\;\;\;e5} +2f^{ab}\cal{F}_{\a\r}\cal{F}_{\mu\nu} \big)\Big\} \, .
\end{align}

\noindent After the symmetry breaking, the $\widehat{S}^{(1)}_{Aff}$ part of the action becomes:
\begin{equation}
\widehat{S}^{(1)}_{Aff} = \frac{1}{16}\theta^{\a\b}\int {\rm d}^{4}x\;e\;\mathcal{F}_{\a\b}f^{2} +h.c. = \frac{1}{8}\theta^{\a\b}\int d^{4}x\;e\;\mathcal{F}_{\a\b}(f^{ab}f_{ab}+2f^{a5}f_{a}^{\;\;5}) \, . \label{NCSAff1}
\end{equation}
The gravitational part (that which includes quantities like curvature and torsion) of $\widehat{S}^{(1)}_{Aff}$ is purely imaginary and it vanishes after adding its hermitian conjugate and so we are left with (\ref{NCSAff1}).

\noindent Now we need to evaluate the action {\small $\widehat{S}_{A}=\widehat{S}_{A}^{(0)}+\widehat{S}_{A}^{(1)}$}
on the equations of motion (EOMs) of the auxiliary field $f$, up to first order in the NC parameter $\theta^{\a\b}$. The EOMs are obtained by varying $\widehat{S}_{A}$ in $f_{ab}$ and $f_{a5}$ independently and we calculate EOMs up to first order in the NC parameter $\theta^{\alpha\beta}$. The first order action, evaluated on the EOMs of the field $f$, denoted as $\widehat{S}^{(1)}_{AEOM}$, has two contributions. The first contribution comes from evaluating the first order action {\small $\widehat{S}_{A}^{(1)}$} on the zeroth order EOMs which have been already calculated in Section 2.1 and are given by (\ref{EOMs-0th order}). The second contribution comes from evaluating the zeroth order action {\small $\widehat{S}_{A}^{(0)}$} on the first order EOMs for the field $f$. It is straightforward, although tedious, to compute the first order EOMs, but actually there is no need for that. It can be readily demonstrated that if we work only up to first order, the zeroth order NC action (\ref{SA}) is annihilated after inserting the first order EOMs for $f$, whatever they may be. This is the consequence of the zeroth order equations (\ref{EOMs-0th order}).  Thus, we only need to insert zeroth order terms (\ref{EOMs-0th order}) in the first order action $\widehat{S}_{A}^{(1)}$. The resulting first order action is a sum
\begin{equation}
\widehat{S}^{(1)}_{AEOM} =\sum_{j=1}^6 \widehat{S}^{(1)}_{AEOMf.j} +  \widehat{S}^{(1)}_{AEOMff}\, .\label{NCSA1EOM} 
\end{equation}
The corresponding terms are given by
\begin{align}
&\widehat{S}^{(1)}_{AEOMf.1} = \frac{1}{32}\theta^{\a\b}\int {\rm d}^{4}x\;e\;
\Big\{ \mathcal{F}^{\m\n}R_{\m\n ab} \big( R_{\a\b}^{\;\;\;ab} -\frac{2}{l^2}e_{\a}^{a}e_{\b}^{b}\big) \nn\\
& + \mathcal{F}^{\r\s} e_{\r}^{a}e_{\s}^{b} \big( R_{\m\n ab}R_{\a\b}^{\;\;\;cd}e_{c}^{\m}e_{d}^{\n}  -\frac{2}{l^2}R_{\a\b ab}\big)
+ 4\mathcal{F}^{\r\m} e_{\r}^{c}\big( R_{\m\n ac}R_{\a\b}^{\;\;\;ab}e_{b}^{\n}  -\frac{2}{l^2}R_{\m\b ac}e_{\a}^{a}\big) \nn\\
& +\mathcal{F}_{\l\t}e^{\l}_{a}e^{\t}_{b}R_{\a\b}^{\;\;\;ab}
\big( R_{\m\n}^{\;\;\;mn}\;e_{m}^{\m}e_{n}^{\n} -\frac{12}{l^2}\big) + \frac{2}{l^2}\mathcal{F}^{\m\n}T_{\a\b}^{\;\;\;a}\big( T_{\m\n a} -2T_{\r\n m}e_{a}^{\r}e_{\m}^{m}\big)\nn\\
& - \frac{2}{l^2}\mathcal{F}_{\a\b}\big( R_{\m\n}^{\;\;\;mn}e_{m}^{\m}e_{n}^{\n} -\frac{12}{l^2} -4l^2\mathcal{F}^{\m\n}\mathcal{F}_{\m\n} \big)\Big\} \, ,\label{1112}
\end{align}
\begin{align}
&\widehat{S}^{(1)}_{AEOMf.2} = \frac{1}{4}\theta^{\a\b}\int {\rm d}^{4}x\;e\;\Big\{
-(\nabla_\b R_{\m\n}^{\;\;\;\;mc})(\nabla_\a e_{\r}^{r})e_{c}^{\l} \big( e_{m}^{\n} (\mathcal{F}_{\l}^{\;\;\m}e_{r}^{\r} - \mathcal{F}_{\l}^{\;\;\r}e_{r}^{\m}) + \mathcal{F}_{\l}^{\;\;\n}e_{r}^{\m}e_{m}^{\r}\big)\nn\\
& + \frac{1}{l^2}\mathcal{F}_{\r}^{\;\;\m}e_{c}^{\r}
\big( \nabla_\b T_{\a\m}^{\;\;\;c} + e_{\b b}R_{\m\n}^{\;\;\;bc} \big) -\frac{4}{l^2}\mathcal{F}_{\n}^{\;\;\m}(\nabla_\a e_{\r}^{r})(\nabla_\b e_\m^m)e_{m}^{\n}e_{r}^{\r} \nn\\
& -\frac{1}{l^2}(\nabla_\a e_{\r}^{r})e_{r}^{\r} \big( e_{c}^{\n}\mathcal{F}_{\b}^{\;\;\m}T_{\m\n}^{\;\;\;\;c} - e_{c}^{\n}\mathcal{F}_{\n}^{\;\;\m}T_{\m\b}^{\;\;\;c} \big)  + \frac{1}{2l^2}T_{\a\b}^{\;\;\;r}T_{\m\n}^{\;\;\;c}\mathcal{F}_{\l}^{\;\;\n}e_{c}^{\l}e_{r}^{\m} \nn\\
& +\frac{1}{l^2}(\nabla_\a e_{\r}^{r})e_{r}^{\n} \Big( T_{\m\n}^{\;\;\;m}\big( \mathcal{F}_{\b}^{\;\;\m}e_{m}^{\r} - \mathcal{F}_{\b}^{\;\;\r}e_{m}^{\m}\big) + T_{\b\n}^{\;\;\;m}\mathcal{F}_{\m}^{\;\;\r}e_{m}^{\m} \Big)\nn\\
& +\frac{2}{l^2} \mathcal{F}_{\l}^{\;\;\r}e_{c}^{\l}(\nabla_\a e_{\r}^{r})(\nabla_\b e_\n^c)e_{r}^{\n} +\frac{1}{l^4}\mathcal{F}_{\a\b}\Big\} \, ,\label{13}
\end{align}
\begin{align}
&\widehat{S}^{(1)}_{AEOMf.3}= -\frac{1}{16}\theta^{\a\b}\int {\rm d}^{4}x\;e\; \Big\{ \mathcal{F}^{\m\n}\Big( R_{\b\n am}\big( R_{\a\m}^{\;\;\;\;am} 
-\frac{4}{l^2}e_\a^a e_\m^m \big)   +8\mathcal{F}_{\a\m}\mathcal{F}_{\b\n} \Big)\nn\\
& +\mathcal{F}_{\l\t}R_{\a\m}^{\;\;\;am}R_{\b\n}^{\;\;\;bn} \big( e_{a}^{\m}e_{m}^{\n}e_{b}^{\l}e_{n}^{\t} + e_{a}^{\l}e_{m}^{\t}e_{b}^{\m}e_{n}^{\n} + 2e_{n}^{\l}e_{m}^{\t}( e_{a}^{\m}e_{b}^{\n} - e_{a}^{\n}e_{b}^{\m})\big)\nn\\
& + \frac{2}{l^2}e_{n}^{\l}e_{b}^{\t} \big( 2\mathcal{F}_{\a\t}R_{\b\l}^{\;\;\;bn} -\mathcal{F}_{\l\t}R_{\a\b}^{\;\;\;bn}\big) 
+ \frac{2}{l^2}\mathcal{F}^{\m\n}T_{\a\m}^{\;\;\;a}T_{\b\n a}\Big\} \, ,\label{14}
\end{align}
\begin{align}
&\widehat{S}^{(1)}_{AEOMf.4} + \widehat{S}^{(1)}_{AEOMf.5} = \frac{1}{16}\theta^{\a\b}\int {\rm d}^{4}x\;e\;\Big\{\nn\\
& + R_{\mu\nu}^{\;\;\;ab}(\nabla_\alpha e_{\rho}^{m})(\nabla_\beta e_{\sigma m})\Big( \mathcal{F}^{\m\n}e_{a}^\r e_{b}^\s 
+ \mathcal{F}^{\r\s}e_{a}^\m e_{b}^\n -4\mathcal{F}^{\m\r}e^{\n}_a e^{\s}_b\Big)\nn\\
&-\frac{1}{l^2}R_{\mu\nu}^{\;\;\;ab}\big( \mathcal{F}^{\m\n}e_{\a a} e_{\b b} 
+ \mathcal{F}_{\a\b}e_{a}^\m e_{b}^\n + 4\mathcal{F}^\m_{\;\;\b}e^{\n}_a e^{\s}_b\big) -\frac{1}{l^2}  \mathcal{F}^{\r\s}(\nabla_\alpha e_{\rho}^{m})(\nabla_\beta e_{\sigma m})\nn\\
& +\frac{2}{l^2}T_{\mu\nu}^{\;\;\;c}(\nabla_\alpha e_\rho^d)\Big( \mathcal{F}^{\m\n}(2e^{\r}_c e_{\b d}  - e^{\r}_d e_{\b c}) 
+ 2 \mathcal{F}_\b^{\;\;\m}(e^{\r}_c e_d^\n  - e^{\r}_d e_c^\n)\nn\\
& +2\mathcal{F}^{\r\m}(2e^{\n}_c e_{\b d}  - e^{\n}_d e_{\b c}) +2\mathcal{F}^\r_{\;\;\b}e^{\m}_c e_d^\n\Big)  
-\frac{2}{l^2}\mathcal{F}^{\r\m}T_{\mu\nu c}T_{\a\b}^{\;\;\;d}e_{\r}^c e_d^\n \label{1516}\\
& -\frac{4}{l^2}T_{\beta\nu a}(\nabla_\alpha e_\rho^d)e^{a}_\l\big( \mathcal{F}^{\l\n}e_d^\r
-\mathcal{F}^{\l\r}e_d^\n \big) + \frac{4}{l^2} R_{\beta\nu\;\;a}^{\;\;\; c}e^{a}_\l\big( \mathcal{F}^{\l\n}e_{\a c}
-\mathcal{F}^\l_{\;\;\a}e_c^\n \big)   -\frac{6}{l^4}\mathcal{F}_{\a\b} \Big\} \, , \nn  
\end{align}
\begin{align}
&\widehat{S}^{(1)}_{AEOMf.6} = -\frac{1}{16}\theta^{\a\b}\int {\rm d}^{4}x\;e\;\Big\{ \mathcal{F}^{\kappa\xi}e_{\kappa}^{a}e_{\xi}^{b}e^{\m}_e e^{\n}_f \big( R_{\a\b}^{\;\;\;ef}R_{\mu\nu ab} -2R_{\b\n}^{\;\;\;ef}R_{\a\m ab} \big) \nn\\
&+ 8\mathcal{F}^{\m\n}\big( {\cal{F}}_{\a\b}{\cal{F}}_{\mu\nu} -2{\cal{F}}_{\a\m}{\cal{F}}_{\b\n}\big) 
+\frac{2}{l^2}\mathcal{F}^{\m\n}e_{\m}^{a}e_{\n}^{b}R_{\a\b ab} -\frac{4}{l^2} \mathcal{F}_{\b\n}\e^{\n}_a e^{\r}_b R_{\alpha\rho}^{\;\;\;ab}  \nn\\
& +\frac{4}{l^2}\mathcal{F}^{\kappa\m}e_{\kappa}^{a}  e^{\r}_e \big( T_{\r\m a}T_{\a\b}^{\;\;\;e}
+T_{\m\b a}T_{\a\r}^{\;\;\;e} + T_{\b\r a}T_{\a\m}^{\;\;\;e}\big) - \frac{8}{l^4}\mathcal{F}_{\a\b}\Big\} \, ,
\label{1718}
\end{align}       
\begin{equation}
\widehat{S}^{(1)}_{AEOMff} = \frac{1}{8}\theta^{\a\b}\int {\rm d}^{4}x\;e\;\mathcal{F}_{\a\b}\mathcal{F}^{\m\n}\mathcal{F}_{\m\n} \, . 
\end{equation}


\subsection{NC Dirac fermion}

The non-interacting Dirac field has already been treated in the framework of $SO(2,3)_\star$ model  in \cite{VojaGocanin2017} where we proposed an NC action for Dirac spinor coupled with gravity. It is the canonically deformed version of the classical action given by the sum of (\ref{Spsi-kin}) and (\ref{Spsi-mass})  
\begin{align}
&\widehat{S}_{\psi} = \widehat{S}_{\psi, kin} + \widehat{S}_{\psi, m} \label{NCDirac}\\
&= \frac{i}{12}\int{\rm d}^{4}x\;\epsilon^{\mu\nu\rho\sigma}\;\Big\{ \widehat{\bar{\psi}}
\star(D_{\mu}\widehat{\phi})\star(D_{\nu}\widehat{\phi}
)\star(D_{\rho}\widehat{\phi})\star(D_{\sigma}\widehat{\psi}) \nn\\
&- (D_{\sigma}\widehat{\bar{\psi}})\star(D_
{ \mu } \widehat{\phi} )\star 
(D_{\nu}\widehat{\phi})\star(D_{\rho}\widehat{\phi}
)\star\widehat{\psi}\Big\}\nn\\
& +\frac{i}{144}\Big(\frac{m}{l}
-\frac{2}{l^{2}}\Big)\int
{\rm d}^{4}x\;\epsilon^{\mu\nu\rho\sigma}\widehat{\bar{\psi}}\star\Big\{
D_{\mu}\widehat{\phi}\star D_{\nu}\widehat{\phi} \star D_{\rho}
\widehat{\phi}\star D_{\sigma}\widehat{\phi}\star\widehat{\phi} \nn\\
&-D_{\mu}\widehat{\phi}\star D_{\nu}\widehat{\phi}\star
D_{\rho}\widehat{\phi}\star\widehat{\phi}\star D_{\sigma}\widehat{\phi} + D_{\mu}\widehat{\phi}\star 
D_{\nu}\widehat{\phi}\star\widehat{\phi}\star D_{\rho}\widehat{\phi}\star 
D_{\sigma}\widehat{\phi}\Big\}\star\widehat{\psi} +h.c. \, .\nn
\end{align}
\noindent Expanding this action via the SW map leads to a non-vanishing first order NC correction after the symmetry breaking. To include interaction, we generalize the result in \cite{VojaGocanin2017} by making substitutions $\nabla_\m\rightarrow\widetilde{\nabla}_\m=\nabla_\m-iA_{\m}$ and $F_{\m\n}\rightarrow\mathbb{F}_{\m\n}=F_{\m\n}+\mathcal{F}_{\m\n}$. 
The fermionic mass term is also modified due to inclusion of the $U(1)$ gauge field. Beside the generalization of the covariant derivative, an additional term of the type $\bar{\psi}\mathcal{F}_{\a\b}\psi$ appears  
\begin{align} 
&\widehat{S}^{(1)}_{\psi, m} =\frac{\theta^{\alpha\beta}}{4}\mass\int {\rm d}^{4}x\;e\;\bar{\psi}\;\Big\{
-i(\nabla_{\alpha}e_{\mu}^{a})e^{\mu}_{a}\widetilde{\nabla}_{\beta} +\frac{1}{6}\eta_{ab}(\nabla_{\alpha}e_{\mu}^{a})(\nabla_{\beta}e_{\nu}^{b})\sigma^{\mu\nu} \nn \\
&-\frac{1}{3}(\nabla_{\alpha}e_{\mu}^{a})(\nabla_{\beta}e_{\nu}^{b})(e^{\mu}_{a}e^{\nu}_{c}-e^{\mu}_{c}e^{\nu}_{a})
\sigma^{c}_{\;\;b} -\frac{1}{9l}(\nabla_{\alpha}e_{\mu}^{a})e^{\mu}_{a}\gamma_{\beta}
-\frac{1}{24}R_{\alpha\beta}^{\;\;\;\;ab}\sigma_{ab} \nn\\
&-\frac{1}{3}R_{\alpha\mu}^{\;\;\;\;ab}e^{\mu}_{a}e_{\beta}^{c}\sigma_{bc} -\frac{1}{18l}T_{\alpha\beta}^{\;\;\;\;a}\gamma_{a} -\frac{7}{18l}T_{\alpha\mu}^{\;\;\;\;a}e^{\mu}_{a}
\gamma_{\beta} -\frac{1}{2l^2}\sigma_{\alpha\beta} -\frac{3}{2}\mathcal{F}_{\a\b} \Big\}\; \psi +h.c. \, . \label{total 
mass term after SB} 
\end{align}
\noindent The complete result for the first order NC correction to the kinetic part after the substitution (i.e. inclusion of $U(1)$ gauge field) is given by:

\begin{align}\label{kinetic-term-afterSB}
& \widehat{S}^{(1)}_{\psi, kin} = \frac{1}{8}\theta^{\a\b}\int d^{4}x\;e\;\bar{\psi}\nn\\
&\Bigg\{ -R_{\alpha\mu}^{\;\;\;\;ab}\Big(e^{\mu}_{a}\gamma_{b}
   +\frac{i}{2}e^{\mu}_{c}\epsilon^{c}_{\;abd}\;\gamma^{d}\gamma^{5}\Big)\widetilde{\nabla}_{\b}
      +\frac{1}{2}R_{\alpha\beta}^{\;\;\;\;ab}\Big(e^{\sigma}_{b}\gamma_{a}-\frac{i}{2} 
 \epsilon_{abc}^{\;\;\;\;\;d}e^{\sigma}_{d}\;\gamma^{c}\gamma^{5}\Big)\widetilde{\nabla}_{
\s} \nn\\
& -\frac{i}{3}R_{\alpha\mu}^{\;\;\;\;ab}
   \epsilon_{abc}^{\;\;\;\;\;d}e^{c}_{\beta}
 (e^{\mu}_{d}e^{\sigma}_{s}-e^{\mu}_{s}e^{\sigma}_{d})\;\gamma^{s}\gamma^{5}\widetilde{
\nabla}_{\s}+\frac{i}{l}T_{\alpha\mu}^{\;\;\;\;a}\Big(e^{\mu}_{a}-i e^{\mu}_{b}\sigma_{a}^{\;\;b}\Big)\widetilde{\nabla}_{\b}\nn\\
& -\frac{i}{l}T_{\alpha\beta}^{\;\;\;\;a}\Big(e^{\sigma}_{a}+\frac{i}{2}e^{\mu}_{a}\sigma_{\mu}^{\;\;\sigma}\Big)\widetilde{\nabla}_{\s}  
   -\frac{2}{3l}
T_{\alpha\mu}^{\;\;\;\;a}\Big( \epsilon_{ab}^{\;\;\;cd}e_{\beta}^{b}e^{\mu}_{c}e^{\sigma}_{d}\gamma^{5}\widetilde{\nabla}_{\s}
+\frac{3}{4l}e^{\mu}_{a}\gamma_{\beta} -e_{\beta a}\gamma^{\mu}\Big)\nn\\
&+\frac{7i}{6l^{2}}\epsilon_{abc}^{\;\;\;\;\;\;d}e^{a}_{\alpha}e^{b}_{\beta}e^{\sigma}_{d}
\gamma^{c}\gamma^{5}\widetilde{\nabla}_{\s}
   -2\Big(
\nea(e^{\mu}_{a}e^{\sigma}_{b}-e^{\sigma}_{a}e^{\mu}_{b})\gamma^{b}
   +\frac{1}{l}\sigma_{\alpha}^{\;\;\sigma}\Big)\widetilde{\nabla}_{\beta}
   \widetilde{\nabla}_{\sigma}\nn\\
&+i\nea\neb e^{\mu}_{c}e^{\nu}_{d}e^{\sigma}_{s}\Big(\frac{2}{3}\epsilon_{b}^{\;\;cds}\gamma_{a}-\eta_{ab}\epsilon^{cdrs}\gamma_{r}\Big)
   \gamma_{5}\widetilde{\nabla}_{\s}
-\frac{2}{3l}e_{\alpha}^{c}(\nabla_{\beta}e^{b}_{\nu})
   \epsilon_{bc}^{\;\;\;ds}e^{\nu}_{d}e^{\sigma}_{s}\gamma_{5}
   \widetilde{\nabla}_{\s}\nn\\
&-\frac{1}{l}\nea(e^{\mu}_{a}e^{\sigma}_{b}-e^{\sigma}_{a}e^{\mu}_{b})e^{c}_{\beta}
\sigma^{b}_{\;\;c}\widetilde{\nabla}_{\s}-\frac{1}{l}\nea \Big(4i e_{a}^{\mu}+e^{\mu}_{b}\sigma_{a}^{\;\;b}\Big)\widetilde{\nabla}_{\b} -\frac{8}{3l^{3}}\sigma_{\alpha\beta}
\nn\\
&+\frac{1}{16l}R_{\alpha\beta}^{\;\;\;\;ab}\sigma_{ab}
   -\frac{1}{2l}
R_{\alpha\mu}^{\;\;\;\;ab}\Big(\frac{5}{3}e^{\mu}_{a}e^{c}_{\beta}+e_{\beta
a}e^{\mu c}\Big)\sigma_{bc}
   -\frac{3}{4l^{2}}T_{\alpha\beta}^{\;\;\;\;a}\gamma_{a} \nn\\ 
&+\frac{2}{3l}\nea\neb\Big(\eta_{ab}\sigma^{\mu\nu}
         -2(e^{\mu}_{a}e^{\nu}_{c}-e^{\mu}_{c}e^{\nu}_{a})\sigma^{c}_{\;\;b}\Big)   
   -\frac{1}{2l^{2}}\nea \Big(3e^{\mu}_{a}\gamma_{\beta}
    -e_{\beta a}\gamma^{\mu}\Big) \nn\\
&+3i\mathcal{F}_{\a\b}e^{\s}_s\gamma^s\widetilde{\nabla}_\s 
-2i\mathcal{F}_{\a\m}e^{\m}_m\gamma^m\widetilde{\nabla}_\b-\frac{5}{l}
\mathcal{F}_{\a\b} \Bigg\} \psi + h.c.\, .
\end{align}
\noindent Apart from the change in covariant derivative, the last three terms in the action, those that include the electromagnetic field strength $\mathcal{F}_{\m\n}$, are completely new. It is important that all three of them pertain in the limit of flat space-time and this leads to new phenomenological consequences concerning NC electrodynamics. Putting the pieces together, we come to the action for NC electrodynamics in curved space-time up to the first order in deformation parameter. It is given by
\bea 
\widehat{S}=\widehat{S}^{(0)}+\widehat{S}^{(1)}_{\psi, kin}+\widehat{S}^{(1)}_{\psi, m}+\widehat{S}^{(1)}_{AEOMf} +\widehat{S}^{(1)}_{AEOMff} \, .
\eea 
\noindent This action is hermitian and invariant under local $SO(1,3)\otimes U(1)$ transformations. From it we can derive the Dirac equation for an electron and the Maxwell's electromagnetic field equations in curved noncommutative space-time. Although this calculation is straightforward, we will not peruse it now. Instead we will investigate in detail the flat NC space-time limit.

\section{Flat space-time NC Electrodynamics in $SO(2,3)_\star$ model}

Although the flat space-time limit might seem too much a simplification, nevertheless in this section we will show that interesting and non-trivial results can be obtained.

\noindent The action for NC electrodynamics in flat space-time up to the first order in $\theta^{\a\b}$ is given by
\begin{align}
\widehat{S}_{flat} =& \widehat{S}^{(0)}_{flat}+\widehat{S}^{(1)}_{flat} \nn\\
=& \int {\rm d}^{4}x\;\bar{\psi}(i\slash{\mathcal{D}} - m)\psi-\frac{1}{4}\int d^{4}x\;\mathcal{F}_{\m\n}\mathcal{F}^{\m\n}\nn\\
&+\theta^{\a\b}\int {\rm d}^{4}x\;\Big(\frac{1}{2}\mathcal{F}_{\a\m}\mathcal{F}_{\b\n}\mathcal{F}^{\m\n}-\frac{1}{8}\mathcal{F}_{\a\b}\mathcal{F}^{\m\n}\mathcal{F}_{\m\n} 
\Big) \nn\\
&+\theta^{\a\b}\int {\rm d}^{4}x\;\bar{\psi}\Bigg(-\frac{1}{2l}
\sigma_{\a}^{\;\;\s}\mathcal{D}_{\b}\mathcal{D}_{\s}
+\frac{7i}{24l^{2}}\epsilon_{\a\b}^{\;\;\;\;\r\s}
\gamma_{\r}\gamma_{5}\mathcal{D}_{\s}  \nn\\
& - \left(\frac{m}{4l^{2}}+\frac{1}{6l^{3}}\right)\sigma_{\alpha\beta} +\frac{3i}{4}\mathcal{F}_{\a\b}\slash{\mathcal{D}}
-\frac{i}{2}\mathcal{F}_{\a\m}\gamma^\m\mathcal{D}_\b
-\left(\frac{3m}{4}-\frac{1}{4l}\right)\mathcal{F}_{\a\b}\Bigg)\psi \, , \label{SQED-flat}
\end{align}
where we introduced flat space-time covariant derivative $\mathcal{D}_{\m}=\partial_{\m}-iA_{\m}$. We notice immediately that this action is different from actions for NC Electrodynamics already present in the literature \cite{minimalNC-Eld}. The new terms appear as residual from the gravitational interaction and they will lead to some non-trivial phenomena, like a deformed dispersion relation and deformed propagator for fermions. Also, we see the appearance of new interaction terms between fermions and the electromagnetic field specific to the $SO(2,3)_{\star}$ model.

\subsection{Deformed equations of motion}

By varying with respect to $A_{\r}$ we obtain NC Maxwell equation with sources in flat space-time. Up to first order the equation is given by
\begin{align}
&\partial_{\m}\mathcal{F}^{\m\r}-\frac{1}{4}\theta^{\a\b}\mathcal{F}_{\a\b}\partial_{\m}\mathcal{F}^{\m\r}
-\frac{1}{2}\theta^{\a\r}\mathcal{F}_{\a\n}\partial_{\m}\mathcal{F}^{\m\n} 
+\theta^{\a\b}\partial_{\m}(\mathcal{F}_{\a}^{\;\;\m}\mathcal{F}_{\b}^{\;\;\r})\label{NCGaugeEOM}\\
&=-\bar{\psi}\gamma^{\r}\psi-\frac{i}{2l}\theta^{\a\r}\bar{\psi}\sigma_{\a}^{\;\;\s}\mathcal{D}_{\s}\psi
-\frac{i}{2l}\theta^{\a\b}\bar{\psi}\sigma_{\a}^{\;\;\r}\mathcal{D}_{\b}\psi
+\frac{i}{2l}\theta^{\a\b}\partial_{\b}(\bar{\psi}\sigma_{\a}^{\;\;\r}\psi) \nn\\
&-\frac{7}{24l^{2}}\theta^{\a\b}\epsilon_{\a\b}^{\;\;\;\;\l\r}\bar{\psi}\gamma_{\l}\gamma_{5}\psi 
-\frac{i}{2}\theta^{\a\b}\partial_{\a}(\bar{\psi}\gamma^{\r}\mathcal{D}_{\b}\psi)
+\frac{i}{2}\theta^{\r\b}\partial_{\m}(\bar{\psi}\gamma^{\m}\mathcal{D}_{\b}\psi)
+\frac{1}{2l}\theta^{\a\r}\partial_{\a}(\bar{\psi}\psi) \, . \nn
\end{align}
By varying NC action (\ref{SQED-flat}) with respect to $\bar{\psi}$ we obtain a deformed Dirac equation for an electron coupled to the electromagnetic field $A_{\m}$
\begin{equation}
\left(i\slash{\partial}-m+\slash{A} 
+\theta^{\a\b}\mathcal{M}_{\a\b}\right)\psi=0 \, , \label{EM-Dirac}
\end{equation}
where $\theta^{\a\b}\mathcal{M}_{\a\b}$ is given by
\begin{align}
\theta^{\a\b}\mathcal{M}_{\a\b}=\theta^{\a\b}\Bigg\{&-\dfrac{1}{2l}\sigma_\a^{\;\;\s}\mathcal{D}_\b \mathcal{D}_\s +\dfrac{7i}{24l^{2}}\epsilon_{\a\b}^{\;\;\;\;\r\s}\gamma_\r\gamma_5 \mathcal{D}_\s- \left(\frac{m}{4l^{2}}+\frac{1}{6l^{3}}\right)\sigma_{\a\b} 
\nn\\
&+\frac{3i}{4}\mathcal{F}_{\a\b}\slash{\mathcal{D}}-
\frac{i}{2}\mathcal{F}_{\a\m}\gamma^\m \mathcal{D}_\b-\left(\frac{3m}{4}-\frac{1}{4l}\right)\mathcal{F}_{\a\b}\Bigg\} \, . \label{thetaM}
\end{align}
From (\ref{thetaM}) we immediately see that there will be new interaction terms in (\ref{EM-Dirac}). Remember that for electron $q=-1$.

\section{Electron in background magnetic field}

We will use the deformed Dirac equation (\ref{EM-Dirac}) with (\ref{thetaM}) to  investigate the special case of an electron propagating in constant magnetic field $\textbf{B}=B\textbf{e}_{z}$. We choose the gauge $A_{\m}=(0, By, 0, 0)$ accordingly. Then, an appropriate ansatz \cite{VojaZbirka} for (\ref{EM-Dirac}) is  
\be 
\psi=\begin{pmatrix}
  \varphi(y)   \\[6pt]
  \chi(y)       
     \end{pmatrix}e^{-iEt+ip_{x}x+ip_{z}z} \, . \label{ansatz}
\ee
Spinor components and energy function are all represented as perturbation series in powers of the deformation parameter,
\bea 
\varphi &=&\varphi^{(0)}+\varphi^{(1)}+\mathcal{O}(\theta^{2})\ , \\
\chi &=&\chi^{(0)}+\chi^{(1)}+\mathcal{O}(\theta^{2})\ , \\
E&=&E^{(0)}+E^{(1)}+\mathcal{O}(\theta^{2}) \, .
\eea 
Inserting the ansatz (\ref{ansatz}) in the Dirac equation (\ref{EM-Dirac}) we obtain
\be 
\left[E\gamma^{0}-p_{x}\gamma^{1}+i\gamma^{2}\frac{d}{dy}-p_{z}\gamma^{3}-m +By\gamma^{1}+\theta^{\a\b}\mathcal{M}_{\a\b}\right]
\begin{pmatrix}
  \varphi(y)   \\[6pt]
  \chi(y)       
     \end{pmatrix}=0 \, .
\ee
The zeroth order (undeformed) equation is given by 
\be 
\left[E^{(0)}\gamma^{0}-p_{x}\gamma^{1}+i\gamma^{2}\frac{d}{dy}-p_{z}\gamma^{3}-m+By\gamma^{1}\right]
\begin{pmatrix}
  \varphi^{(0)}   \\[6pt]
  \chi^{(0)}      
     \end{pmatrix}=0 \, , \label{zeroth-order}
\ee
while the first order equation is
\bea 
&&\left[E^{(0)}\gamma^{0}-p_{x}\gamma^{1}+i\gamma^{2}\frac{d}{dy}-p_{z}\gamma^{3}-m+By\gamma^{1}\right]
\begin{pmatrix}
  \varphi^{(1)}   \\[6pt]
  \chi^{(1)}       
     \end{pmatrix}\nn\\
&&\;\;\;\;\;\;\;=-\left[E^{(1)}\gamma^{0}+\theta^{\a\b}\mathcal{M}_{\a\b}\right]
\begin{pmatrix}
  \varphi^{(0)}   \\[6pt]
  \chi^{(0)}     
\end{pmatrix} \, .\label{first-order}
\eea
Bearing in mind that the adjoint of (\ref{zeroth-order}) is 
\be 
\bar{\psi}^{(0)}\left[E^{(0)}\gamma^{0}-p_{x}\gamma^{1}-i\gamma^{2}\overleftarrow{\frac{d}{dy}}-p_{z}\gamma^{3}-m+By\gamma^{1}\right]=0 \, ,\label{adjoint-equation}
\ee
after multiplying (\ref{first-order}) by $\bar{\psi}^{(0)}$ from the left and integrating over $y$ we obtain:
\begin{equation}
E^{(1)}\int {\rm d}y\;\bar\psi^{(0)}\g^{0}\psi^{(0)}=-\theta^{\a\b}\int {\rm d}y\;\bar\psi^{(0)}{\cal M}_{\a\b}\psi^{(0)} \, .  \nn
\end{equation}
Therefore, the NC energy correction can be calculated as
\begin{equation}
E^{(1)} = \frac{-\theta^{\a\b}\int {\rm d}y\;\bar\psi^{(0)}{\cal M}_{\a\b}\psi^{(0)}}{\int {\rm d}y\;\bar\psi^{(0)}\g^{0}\psi^{(0)}} \, .\label{energy-shift}
\end{equation}

\noindent Let us calculate explicitly the zeroth order solution $\psi^{(0)}$. From the unperturbed equation (\ref{zeroth-order}) we can derive the equation for $\varphi^{(0)}$ spinor component. It is given by
\be 
\left[\frac{{\rm d}^{2}}{{\rm d}y^{2}}-(p_{x}-By)^{2}+(E^{(0)})^{2}-p_{z}^{2}-m^{2}-B\sigma_{3}\right]\varphi^{(0)}(y)=0 \, .\nn
\ee
It is well known that the unperturbed energy levels of an electron in constant magnetic field (relativistic Landau levels) are
\be 
E_{n,s}^{(0)}=\sqrt{p_{z}^{2}+m^{2}+(2n+s+1)B} \, , \label{E-undeformed}
\ee
where $n$ is the principal quantum number and $n=0,1,2,\dots$, while $s=\pm 1$ are the eigenvalues of the matrix $\sigma_3$. 

\noindent The complete undeformed Dirac spinor is
\be 
\psi^{(0)}_{n,s}=\begin{pmatrix}
\varphi^{(0)}_{n,s} \\[6pt]
\chi^{(0)}_{n,s}
\end{pmatrix}e^{-iE_{n,s}^{(0)}t+ip_{x}x+ip_{z}z} \, , \label{total-spinor}
\ee
with components
\begin{align}
& \varphi^{(0)}_{n,s} = \Phi_{n}\varphi_{s}\ , \\
& \chi^{(0)}_{n,s} = \frac{1}{E_{n,s}^{(0)}+m}\Big[-\sqrt{\frac{B}{2}}\big( \sqrt{n+1}\Phi_{n+1}\sigma_{-}
-\sqrt{n}\Phi_{n-1}\sigma_{+}\big) + p_{z}\Phi_{n}\sigma_{3}\Big]\varphi_{s} \, , 
\end{align}
where $\sigma_{\pm}=\sigma_{1}\pm i\sigma_{2}$, and $\varphi_{s}$ is the eigenvector of $\sigma_{3}$ for eigenvalue $s=\pm 1$. Functions $\Phi_{n}(\xi)$ $(n=0,1,...)$ are Hermitian functions defined by
\be 
\Phi_{n}(\xi)=\frac{1}{\sqrt{2^{n}n!\sqrt{\pi B}}}H_{n}\left(\frac{\xi}{\sqrt{B}}\right)e^{-\frac{\xi^{2}}{2B}} \, ,\nn
\ee   
where $H_{n}$ are Hermitian polynomials and $\xi=By-p_{x}$.

\noindent Normalization for (\ref{total-spinor}) gives us 
\begin{align}
\int dy\; \bar{\psi}^{(0)}_{n,s}\gamma^{0}\psi_{n,s}^{(0)}=\frac{2E_{n,s}^{(0)}}{B(E_{n,s}^{(0)}+m)} \, .
\end{align}
We are looking for NC shift $E^{(1)}_{n,s}$ of the ordinary, undeformed energy levels (\ref{E-undeformed}). Using the equation $(\ref{energy-shift})$, we find
\begin{equation}
E^{(1)}_{n,s} = -\frac{\theta^{\a\b}\int dy\;\bar\psi^{(0)}_{n,s}{\cal M}_{\a\b}\psi^{(0)}_{n,s}}{\int dy\;\bar\psi^{(0)}_{n,s}\g^{0}\psi^{(0)}_{n,s}} \, . \label{EnergyCorrectionExpl}
\end{equation}
In particular, for $\theta^{12}=-\theta^{21}=\theta\neq 0$ and all the other components of the matrix $\theta^{\a\b}$ equal to zero, we find
\begin{equation}
E^{(1)}_{n,s} = -\frac{\theta s}{E^{(0)}_{n,s}}\left(\frac{m^{2}}{12l^{2}}-\frac{m}{3l^{3}}\right)\Big( 1 + \frac{B}{(E^{(0)}_{n,s}+m)}(2n+s+1)\Big) + \frac{\theta B^{2}}{2E^{(0)}_{n,s}}(2n+s+1) \, . \label{NC-energy-correction}
\end{equation}

\noindent In the absence of magnetic field we confirm the already established result \cite{VojaGocanin2017},
\begin{equation}
E^{(1)}_{n,s}=-\frac{\theta s}{E^{(0)}_{n,s}}\left( \frac{m^{2}}{12l^{2}}-\frac{m}{3l^{3}}\right) \, .\nn
\end{equation}
The NC energy levels depend on $s=\pm 1$ and we see that constant noncommutative background causes Zeeman-like splitting of undeformed energy levels.

\noindent The non-relativistic limit of NC energy levels (\ref{NC-energy-correction}) is obtained by expanding the undeformed energy function $E^{(0)}_{n,s}$ assuming that $p_{z}^{2}$, $B\ll m^{2}$:
\bea 
E_{n,s}^{(0)} &=& m+\frac{p_{z}^{2}+(2n+s+1)B}{2m}-\frac{(p^{2}_{z}+(2n+s+1)B)^{2}}{8m^{3}} \, .  \nn
\eea
Expanding (\ref{NC-energy-correction}), we obtain the first order NC correction to the energy levels of a non-relativistic electron:
\bea
E^{(1)}_{n,s} &=& \left(\frac{\theta s}{3l^{3}}-\frac{\theta sm}{12l^{2}}\right)\Big(1-\frac{p^{2}_{z}}{2m^{2}}+\frac{3p_{z}^{4}}{8m^{4}}+\frac{3p_{z}^{2}(2n+s+1)B}{8m^{4}}\Big)\nn\\
&& +\frac{\theta B^{2}}{2m}(2n+s+1)\Big(1-\frac{p_{z}^{2}+(2n+s+1)B}{2m^{2}}+\frac{3(p_{z}+(2n+s+1)B)^{2}}{8m^{4}}\Big) \, .\nn
\eea
If we take $p_{z}=0$, the non-relativistic NC energy levels reduce to
\begin{align}
&E_{n,s} = E_{n,s}^{(0)}+E_{n,s}^{(1)}+\mathcal{O}(\theta^{2}) \nn\\
&= m +\frac{\theta s}{3l^{3}}-\frac{\theta s m}{12l^{2}} + \frac{2n+s+1}{2m}(B+\theta B^{2})-\frac{(2n+s+1)^{2}}{8m^{3}}(B^{2}+2\theta B^{3})+\mathcal{O}(\theta^{2})\nn\\
&= m-s\theta\left(\frac{m}{12l^{2}}-\frac{1}{3l^{3}}\right) + \frac{2n+s+1}{2m}B_{eff}-\frac{(2n+s+1)^{2}}{8m^{3}}B_{eff}^{2}+\mathcal{O}(\theta^{2}) \, , \label{NC-energylevels-B}
\end{align}
where we introduced $B_{eff}=(B+\theta B^{2})$ as an \textit{effective magnetic field}. As in the case of non-interacting electrons \cite{VojaGocanin2017}, the spin-dependent shift of mass is apparent. If we compare this expression with the one for undeformed energy levels $E_{n,s}^{(0)}$, we see that the only effect of noncommutaivity is to modify the mass of an electron and the value of the background magnetic field. This interpretation of constant noncommutativity is in accord with string theory. In \cite{SWMapEnvAlgebra} it is argued that coordinate functions of the endpoints of an open string constrained to a D-brane in the presence of a constant Neveu-Schwarz B-field satisfy constant noncommutativity algebra. The implication is that a relativistic field theory on NC space-time can be interpreted as a low energy limit, i.e. an effective theory, of the theory of open strings. 

\noindent From the energy function (\ref{NC-energylevels-B}) we can derive a NC deformation of the induced magnetic moment of an electron in the $n$-th Landau level in the limit of a weak magnetic field:
\begin{equation}
\mu_{n,s}=-\frac{\partial E_{n,s}}{\partial B}=-\mu_{B}(2n+s+1)(1+\theta B) \, , \label{NCMgMoment}
\end{equation}
where $\mu_{B}=\tfrac{e\hslash}{2mc}$ is the Bohr magneton. We recognise $-(2n+1)\mu_{B}$ as the diamagnetic moment of an electron and $-s\mu_{B}$ as the spin magnetic moment. The $\theta B$-term  is another potentially observable phenomenological prediction of our model. It is a linear NC correction to the induced electron's dipole moment. A natural next step would be to calculate the induced magnetisation of a material as the induced magnetic moment per unit area. To do this we need to understand better the meaning and the realization of noncommutaivity in materials.
 

\section{Discussion}

In this paper we discussed coupling of matter fields with gravity in the framework of NC $SO(2,3)_\star$ gauge theory of gravity. Using the Seiberg-Witten map we constructed the gauge invariant actions and calculated equations of motion for the NC $U(1)_\star$ gauge field and the NC Dirac fermion. Unlike in \cite{VojaGocanin2017}, fermions are now coupled with NC $U(1)$ gauge field and the NC gravity. In this way we formulated NC Electrodynamics in curved space-time induced by NC $SO(2,3)_\star$ gravity. The flat space-time limit of this model enables one to study behaviour of an electron in a background electromagnetic field. Especially, corrections to the relativistic Landau levels of an electron in a constant magnetic field are given by (\ref{NC-energy-correction}) and their non-relativistic limit is (\ref{NC-energylevels-B}). Motion of an electron in a constant background magnetic field and NC corrections to Landau levels were investigated in the case of canonical noncommutaivity in \cite{NCLLevelsCanonical} and for other types of NC space-times in \cite{NCLLevelsOthers}. It can be seen both from (\ref{NC-energy-correction}) and (\ref{NC-energylevels-B}) that NC correction to (non)-relativistic Landau levels depends on the mass $m$, the principal quantum number $n$ and the spin $s$. In particular, the NC correction to energy levels will be different for different levels. It would be interesting to calculate the NC correction to the degeneracy of Landau levels and we plan to address this problem in future work. It is well known that the physics of the Lowest Landau Level (LLL) is closely related to the physics of Quantum Hall Effect (QHE). Using the obtained results, we plan to investigate NC corrections to the QHE. In this way, together with the induced NC magnetic moment (\ref{NCMgMoment}) and the NC-induced magnetization in materials we hope to obtain some constraints on noncommutaivity parameter from condensed matter experiments.

\noindent Starting from (\ref{SQED-flat}) one can check renormalizability of the model. It is known that, the so-called Minimal NC Electrodynamics, a theory obtained by directly introducing NC Moyal-Weyl $\star$-product in the classical Dirac action for fermions coupled with $U(1)$ gauge field in Minkowski space,
\bea
\widehat{S}=\int d^{4}x\;\widehat{\bar{\psi}}\star(i\gamma^{\m}D_{\m}-m)\widehat{\psi}-\frac{1}{4}\int d^{4}x\;\widehat{F}_{\m\n}\star\widehat{F}^{\m\n} \, , \nn
\eea
is not a renormalizabile theory because of the fermionic loop contributions \cite{minimalNC-Eld}. It would be interesting to see if additional terms present in the NC $SO(2,3)_\star$ gravity induced Electrodynamics (\ref{SQED-flat}) can improve this behaviour. 

\noindent The NC $SO(2,3)_\star$ gravity model also enables one to introduce coupling with non-Abelian gauge fields. In this way, it is possible to progress towards generalizing Standard Model to a NC space-time using the setup we described in this paper. All this open problems and research proposals we postpone for future work.

\vskip1cm \noindent 
{\bf Acknowledgement}    
The work is
supported by project 
ON171031 of the Serbian Ministry of Education and Science and partially
supported by the Action MP1405 QSPACE from the Europe
an Cooperation in Science
and Technology (COST).

\appendix

\renewcommand{\theequation}{\Alph{section}.\arabic{equation}}
\initiate
\section{AdS algebra and the $\Gamma$-matrices}

\noindent Algebra relations\footnote{$\epsilon^{01235}=+1,\ \epsilon^{0123}=+1$}:
\bea
&&\{M_{AB},\Gamma_C\}=i\epsilon_{ABCDE}M^{DE}\nn\\
&&\{M_{AB},M_{CD}\}=\frac{i}{2}\epsilon_{ABCDE}\Gamma^{E}+\frac12(\eta_{AC}\eta_
{BD}-\eta_{AD}\eta_{BC})\nn\\
&&{[}M_{AB},\Gamma_C{]}=i(\eta_{BC}\Gamma_A-\eta_{AC}\Gamma_B)\nn\\
&&\Gamma_A^\dagger=-\gamma_0\Gamma_A\gamma_0\nn\\
&&M_{AB}^\dagger=\gamma_0M_{AB}\gamma_0 \label{AdSrelations}
\eea
Identities with traces:
\bea
&&\tr (\Gamma_A\Gamma_B)=4\eta_{AB}\nn\\
&&\tr (\Gamma_A)=\tr (\Gamma_A\Gamma_B\Gamma_C)=0\nn\\
&&\tr
(\Gamma_A\Gamma_B\Gamma_C\Gamma_D)=4(\eta_{AB}\eta_{CD}-\eta_{AC}\eta_{BD}+\eta_
{AD}\eta_{CB})\nn\\
&&\tr (\Gamma_A\Gamma_B\Gamma_C\Gamma_D\Gamma_E)=-4i\epsilon_{ABCDE}\nn\\
&&\tr (M_{AB}M_{CD}\Gamma_E)=i\epsilon_{ABCDE}\nn\\
&&\tr (M_{AB}M_{CD})=-\eta_{AD}\eta_{CB}+\eta_{AC}\eta_{BD} \nn \\ 
&&Tr(M_{AB}\Gamma_{E}\Gamma_{F}\Gamma_{G})=2\varepsilon_{ABEFG} \nn \\
&&Tr(M_{AB}M_{CD}\Gamma_{E}\Gamma_{F}\Gamma_{G})=i\varepsilon_{ABCDE}\eta_{FG}-i\varepsilon_{ABCDF}\eta_{EG}+i\varepsilon_{ABCDG}\eta_{EF} \nn\\
&&\;\;\;\;\;\;\;\;\;\;\;\;\;\;\;\;\;\;\;\;\;\;\;\;\;\;\;\;\;\;\;\;\;\;\;\;\;\;\;+i\varepsilon_{BCEFG}\eta_{AD}+i\varepsilon_{ADEFG}\eta_{BC}\nn\\
&&\;\;\;\;\;\;\;\;\;\;\;\;\;\;\;\;\;\;\;\;\;\;\;\;\;\;\;\;\;\;\;\;\;\;\;\;\;\;\;
-i\varepsilon_{BDEFG}\eta_{AC}-i\varepsilon_{ACEFG}\eta_{BD} \label{AdStraces}
\eea


\begin{thebibliography}{99}

\bibitem{UVIR}
S. Minwalla, M. Van Raamsdonk and N. Seiberg,
{\it Noncommutative perturbative dynamics}, JHEP {\bf 0002}, 020 (2000), [hep-th/9912072];

M.~Van Raamsdonk and N.~Seiberg,
{\it Comments on noncommutative perturbative dynamics}, JHEP {\bf 0003}, 035 (2000), [hep-th/0002186];

R.~J.~Szabo, {\it Quantum field theory on noncommutative spaces},
Phys.\ Rept.\ {\bf 378}, 207 (2003), [hep-th/0109162].

\bibitem{NCSM}
W.~Behr, N.~G.~Deshpande, G.~Duplan\v ci\' c, P.~Schupp, J.~Trampeti\' c and J.~Wess,
{\it The Z $\to$ gamma gamma, gg Decays in the Noncommutative Standard Model},
Eur.\ Phys.\ J. {\bf C29}, 441 (2003) [arXiv:hep-ph/0202121];

B.~Meli\' c, K.~Passek-Kumeri\v cki, P.~Schupp, J.~Trampeti\' c and M.~Wohlgennant,
{\it The Standard Model on Non-Commutative Space-Time: Electroweak
Currents and Higgs Sector}, Eur.\ Phys.\ J. {\bf C42}, 483 (2005)
[arXiv:hep-ph/0502249];

B.~Meli\' c, K.~Passek-Kumeri\v cki, P.~Schupp, J.~Trampeti\' c and M.~Wohlgennant,
{\it The Standard Model on Non-Commutative Space-ime: Strong Interactions
Included}, Eur.\ Phys.\ J. {\bf C42}, 499 (2005)
[arXiv:hep-ph/0503064].

\bibitem{fuzzy}
J. Madore, {\it The Fuzzy Sphere}, Class.\ Quant.\ Grav. {\bf 9}, (1992) 69-88;

M. Buri\' c and J. Madore, {\it Spherically Symmetric Noncommutative Space: $d =
4$}, Eur.\ Phys.\ J. {\bf C58},  347 (2008), [arXiv: 0807.0960].


\bibitem{DynNC}
P.~Aschieri, L.~Castellani and M. Dimitrijevi\' c,
{\it  Dynamical noncommutativity and Noether theorem in twisted
$\Phi^4_\star$ theory}, Lett.\ Math.\ Phys. {\bf 85}, 39 (2008), [arXiv:0803.4325].

P. Aschieri and L. Castellani, {\it Extended gravity theories from dynamical noncommutativity}, Gen.\ Rel.\ Grav.\ {\bf 45}, 411-426 (2013), [arXiv:1206.4096].


\bibitem{SWMapEnvAlgebra}
N.~Seiberg and E.~Witten,
{\it String theory and noncommutative geometry},
JHEP {\bf 09}, 032 (1999), [hep-th/9908142].

\bibitem{NCGrSvi} 
A. H. Chamseeddine, {\it Deforming Einstein's gravity},
Phys.\ Lett.\ B {\bf 504} 33 (2001), [hep-th/0009153].

P.Aschieri, C. Blohmann, M. Dimitrijevi\' c, F. Meyer, P. Schupp
and J. Wess, {\it A Gravity Theory on Noncommutative Spaces},
Class.\ Quant.\ Grav. {\bf 22}, 3511 (2005), [hep-th/0504183].

H. Steinacker, {\it Emergent Geometry and Gravity from Matrix Models: an
Introduction}, Class.\ Quant.\ Grav.\ {\bf 27}, 133001
(2010), [arXiv:1003.4134].

M. Dobrski, {\it On some models of geometric noncommutative general 
relativity}, Phys.\ Rev.\ D {\bf 84}, 065005 (2011), [arXiv:1011.0165].

M. Buri\'c, T. Grammatikopoulos, J. Madore, G. Zoupanos, {\it Gravity and the 
Structure of
Noncommutative Algebras}, JHEP {\bf 0604} 054, 2006, [hep-th/0603044].

\bibitem{Us-16}
M. Dimitrijevi\' c \'Ciri\'c, B. Nikoli\'c and V. Radovanovi\' c, {\it 
NC $SO(2,3)_\star$
gravity: noncommutativity as a source of curvature and torsion}, Phys. Rev. D {\bf 87},  024017 (2017),
[arXiv:1612.00768].
 
\bibitem{MiSO23Razno}
M. Dimitrijevi\' c, V. Radovanovi\' c and H. \v Stefan\v ci\' c,
{\it AdS-inspired noncommutative gravity on the Moyal plane},
Phys. Rev. D {\bf 86}, 105041 (2012), [arXiv:1207.4675].

M. Dimitrijevi\' c and  V. Radovanovi\' c,
{\it Noncommutative $SO(2,3) $ gauge theory and noncommutative gravity},
Phys. Rev. D {\bf 89}, 125021 (2014), [arXiv:1404.4213].

M. Dimitrijevi\' c \'Ciri\'c, B.~Nikoli\' c and V. Radovanovi\' c, {\it Noncommutative gravity and the relevance of the theta-constant deformation},
EPL {\bf 118}, 2 (2017), [arXiv:1609.06469].

\bibitem{VojaGocanin2017}
D. Go\v{c}anin and V. Radovanovi\'c, {\it Dirac field and gravity in NC $SO(2,3)_\star$ model}, Eur. Phys. J. C, \textbf{78} (2018) 195.


\bibitem{stelle-west} K. S. Stelle and P. C. West, {\it Spontaneously
broken de Sitter symmetry and the gravitational holonomy group}, Phys.\ Rev.\ D
{\bf 21}, 1466 (1980).

S. W. MacDowell and F. Mansouri,
{\it Unified geometrical theory of gravity and supergravity}, Phys.\ Rev.\ Lett.
{\bf 38}, 739 (1977).

P. K. Towsend, {\it Small-scale structure of spacetime as
the origin of the gravitation constant}, Phys.\ Rev.\ D {\bf 15},  2795 (1977).

\bibitem{Wilczek}
F. Wilczek, {\it  	
Riemann-Einstein structure from volume and gauge symmetry}, Phys.\ Rev.\ Lett.\ 
{\bf 80} (1998)
4851-4854, [hep-th/9801184].

\bibitem{PLGaugeGr}
P. Aschieri, L. Castellani, {\it Noncommutative gauge fields coupled to noncommutative gravity},  Gen.\ Rel.\ Grav.\ {\bf 45}, 581-598 (2013), [arXiv:1205.1911].

\bibitem{PLGR-fer}
P. Aschieri and L. Castellani, {\it Noncommutative $D=4$ gravity
coupled to fermions} JHEP, {\bf 0906}, 086 (2009), [arXiv:0902.3823].

P. Aschieri and L. Castellani, {\it 	
Noncommutative gravity coupled to fermions: second order expansion via Seiberg-Witten map}, JHEP {\bf 1207} (2012) [arXiv:1111.4822]. 

\bibitem{PLM-13} 
P. Aschieri, L. Castellani and  M.
Dimitrijevi\'c, {\it Noncommutative gravity at second order via 
Seiberg-Witten map}, 
Phys.\ Rev.\ D {\bf 87}, 024017 (2013), [arXiv:1207.4346]. 
 
\bibitem{VojaZbirka} 
V. Radovanovi\'{c}, \textit{Problem Book in Quantum Field Theory}, Springer (2008). 
 
\bibitem{NCLLevelsCanonical}
P. A. Horvathy, {\it The non-commutative Landau problem}, Ann.\ Phys.\ {\bf 299}, 128-140 2002, [hep-th/0201007].

E. V. Gorbar, M. Hashimoto, V.A. Miransky, {\it Nondecoupling phenomena in QED in a magnetic field and noncommutative QED}, Phys.\ Lett.\ B {\bf 611}, (2005) 207-214, [hep-th/0501135].

A. P. Balachandran, K. S. Gupta, S. Kurkcuoglu, {\it Interacting Quantum Topologies and the Quantum Hall Effect},  	Int.\ J.\ Mod.\ Phys.\ A {\bf 23}, 1327-1336 (2008), [arXiv:0707.1219].


\bibitem{NCLLevelsOthers}
R. Iengo, R. Ramachandran, {\it Landau Levels in the noncommutative $AdS_2$}, JHEP {\bf 0202} (2002) 017, [hep-th/0111200]; 

F. M. Andrade, E. O. Silva, D. Assafr\"{a}o, C. Filgueiras, {\it Effects of quantum deformation on the integer quantum Hall effect} EPL \textbf{116}, 31002 (2016), [arXiv:1603.08859].

\bibitem{minimalNC-Eld}
M. Buri\'c and V. Radovanovi\'c, {\it The One loop effective action for 
quantum electrodynamics on noncommutative 
space} JHEP {\bf 0210} (2002) 074, [hep-th/0208204]. 

R. Wulkenhaar, {\it Nonrenormalizability of theta expanded 
noncommutative QED}, JHEP {\bf 0203} (2002) 024, [hep-th/0112248]. 

M. Buri\'c, D. Latas, V. Radovanovi\'c, {\it Renormalizability of 
noncommutative SU(N) gauge theory},  JHEP {\bf 0602} (2006) 046, [hep-th/0510133]. 

T.C. Adorno, D.M. Gitman, A.E. Shabad, D.V. 
Vassilevich, {\it Classical Noncommutative Electrodynamics with External 
Source}, Phys.\ Rev.\ D {\bf 84} (2011) 065003, [arXiv:1106.0639].


\end{thebibliography}
\end{document}